\documentclass[a4paper,11pt]{article}

\usepackage{jheppub} 

\usepackage[T1]{fontenc} 
\usepackage{mathabx}
\usepackage{amsmath}
\usepackage{orcidlink} 

\usepackage{multicol}

\def\beq{\begin{equation}}
\def\eeq{\end{equation}}
\def\bal{\begin{aligned}}
\def\eal{\end{aligned}}

\newcommand{\llangle}{\langle\!\langle}
\newcommand{\rrangle}{\rangle\!\rangle}

\newcommand{\smallarrow}{\mathrel{\scalebox{0.7}[0.7]{$\rightarrow$}}}

\def\p{\prime}

\def\d{{\rm d}}

\def\SL{SL(2,\mathbb R)}
\def\sl{\mathfrak{sl}(2,\mathbb R)}
\def\Vir{{\rm Vir}}
\def\vir{\mathfrak{vir}}

\def\mbms{\mathfrak{maxwell}{\textnormal-}\mathfrak{bms}_3}

\def\bms{\mathfrak{bms}_3}

\title{\boldmath Boundary dynamics of Maxwell-invariant three-dimensional Chern-Simons gravity}

\author[a]{Felix H\"ofenstock
\orcidlink{0009-0004-5243-6161}
}
\author[b,c]{and \;Patricio Salgado-Rebolledo
\orcidlink{0000-0001-6832-6785}
}

\affiliation[a]{Institute for Theoretical Physics, TU Wien, Wiedner Hauptstrasse 8-10/136, A-1040 Vienna, Austria}
\affiliation[b]{Asia Pacific Center for Theoretical Physics (APCTP), Pohang, Gyeongbuk 37673, Korea}
\affiliation[c]{Instituto de Ciencias Exactas y Naturales (ICEN), Universidad Arturo Prat,\\
Playa Brava 3256, 1111346 Iquique, Chile}

\emailAdd{felix.hoefenstock@protonmail.com,salgado.rebolledo@apctp.org}

\abstract{We construct a two-dimensional dual field theory induced at the boundary of three-dimensional Chern-Simons gravity invariant under the Maxwell algebra. The resulting action takes the form of a Maxwellian extension of the flat Liouville theory known from the analysis of asymptotically flat three-dimensional gravity. This boundary theory is derived by reducing the bulk gravitational action to a Maxwell-invariant chiral Wess–Zumino–Witten model and imposing boundary conditions compatible with asymptotically flat geometries. Alternatively, we obtain the same theory as the geometric action on coadjoint orbits of the Maxwell extension of the BMS$_3$ group. Finally, we show how the boundary actions corresponding to both Poincar\'e and Maxwell invariance emerge from a Carrollian expansion of the boundary theory dual to AdS$_3$ Chern-Simons gravity.}

\begin{document}
\maketitle
\flushbottom

\section{Introduction}

Extensions of anti-de Sitter (AdS) gravity and conformal field theory (CFT) have paved the way for generalizing the holographic duality beyond the original AdS/CFT framework \cite{Maldacena:1997re}. In this broader context, several alternative holographic correspondences have been proposed, including dS/CFT \cite{Strominger:2001pn}, Kerr/CFT \cite{Guica:2008mu}, non-relativistic holography \cite{Kachru:2008yh,Balasubramanian:2008dm,Son:2008ye,Bagchi:2010zz}, and flat holography \cite{Barnich:2009se,Barnich:2010eb,Bagchi:2012yk}. In particular, flat holography has recently gained renewed interest in the form of celestial holography \cite{Strominger:2017zoo,Pasterski:2016qvg,Adamo:2019ipt,Donnay:2020guq} and later as Carrollian holography \cite{Donnay:2022aba,Bagchi:2022emh}, both of which explore new formulations of the holographic principle in asymptotically flat spacetimes.

An important playground where these ideas can be explored is three-dimensional gravity. In fact, a precursor to the AdS/CFT correspondence is the seminal result by Brown and Henneaux, who showed that the asymptotic symmetry algebra of asymptotically AdS$_3$ geometries in three spacetime dimensions is given by two copies of the Virasoro algebra, i.e., the centrally extended conformal symmetry in two dimensions \cite{Brown:1986nw}. Key developments following this discovery include Strominger’s derivation of the Bekenstein–Hawking entropy for the BTZ black hole from a purely CFT perspective, using the Brown–Henneaux central charge and the Cardy formula \cite{Strominger:1997eq}; and the identification of the classical boundary dual to AdS$_3$ Einstein gravity with Brown–Henneaux boundary conditions as Liouville theory, by Coussaert, Henneaux, and van Driel \cite{Coussaert:1995zp}.

These results were later generalized to the case of asymptotically flat three-dimensional gravity, where the role of the BTZ black hole is played by a cosmological solution, and suitable boundary conditions lead to an asymptotic symmetry given by the $\bms$ algebra together with a boundary theory described by the flat Liouville model \cite{Barnich:2006av,Barnich:2012xq,Barnich:2013yka}. These results can also be obtained as a vanishing cosmological constant limit or ``flat limit'' of their corresponding AdS counterparts. At the boundary, this is in concordance with the idea of Carrollian holography. Indeed, the flat Liouville model presented in \cite{Barnich:2012rz} is equivalent to the electric Carroll limit of Liouville theory in the sense of \cite{Henneaux:2021yzg}, which finds a natural explanation in the fact that the conformal extension of the Carroll algebra is isomorphic to the BMS symmetry \cite{Duval:2014uva}. Along similar lines, in the context of higher-spin gravity, flat $\mathcal W_N$ algebras have been shown to follow from ultra-relativistic limits of $\mathcal W_N\otimes \mathcal W_N$ algebras \cite{Campoleoni:2016vsh}, characteristic of higher-spin extensions of AdS$_3$ Chern–Simons gravity \cite{Henneaux:2010xg,Campoleoni:2010zq}.

With the aim of extending notions of the AdS/CFT correspondence to asymptotically flat spacetimes, an intriguing question is whether minimal deformations of Poincar\'e symmetry can consistently support a bulk/boundary duality, especially within the tractable framework provided by three-dimensional Chern–Simons gravity. In this context, the Maxwell algebra arises as a natural candidate. It extends the Poincar\'e algebra by including additional antisymmetric tensorial generators, making translations non-Abelian via the relation
\begin{equation}\label{MaxalgD}
[P_a, P_b] = Z_{ab}.
\end{equation}
Historically, this symmetry appeared in the study of particles coupled to a constant electromagnetic field, where magnetic translations include minimal coupling to a gauge potential linear in the coordinates, $P_a = \partial_a -\frac12 Z_{ab}x^b$, and obey a non-vanishing commutator proportional to the field strength \cite{Schrader:1972zd} (see also \cite{Bacry:1970ye,Hoogland:1978wi,Beckers:1983gp,negro1990local}). In the context of gravity, the Maxwell algebra allows one to extend General Relativity by incorporating a generalized cosmological term \cite{deAzcarraga:2010sw,deAzcarraga:2012qj}. In recent years, a wide range of generalized Maxwell-like symmetries have been proposed \cite{Soroka:2004fj,Gomis:2009dm,Gibbons:2009me,Khasanov:2011jr,Salgado:2023owk}, including infinite-dimensional extensions \cite{deAzcarraga:2007et,Bonanos:2008ez,Penafiel:2016ufo,Gomis:2017cmt,Gomis:2020wrv}, some of which have been used to construct gravity models \cite{Edelstein:2006se,Izaurieta:2009hz,Durka:2011va,Salgado:2013eut,Concha:2013uhq,Concha:2014zsa,Gonzalez:2014tta,Aviles:2022tvh}.

Interestingly, to the best of the authors' knowledge, the earliest appearance of Maxwell gravity occurs in the work of Cangemi \cite{Cangemi:1992ri} (see also \cite{Duval:2008tr}), where an extended Poincar\'e algebra was required to obtain a first-order formulation of Jackiw–Teitelboim or CGHS gravity upon dimensional reduction. Moreover, the first-order formulation of CGHS gravity is described by a BF theory invariant under a two-dimensional version of the Maxwell algebra \cite{Cangemi:1992bj,Afshar:2019axx}. The Maxwell-invariant three-dimensional Chern–Simons theory introduced by Cangemi has recently been revisited from various angles. In contrast to Poincar\'e-invariant Chern–Simons gravity, it admits black hole solutions that closely resemble the BTZ black hole \cite{Hoseinzadeh:2014bla}; it features an infinite-dimensional asymptotic symmetry algebra given by an extension of the $\bms$ algebra with non-commuting supertranslations \cite{Concha:2018zeb}; and it has even been proposed as an effective model for the boundary states of a topological insulator \cite{Palumbo:2016nku}. 

Given the fact that the flat Liouville model, dual to Poincar\'e-invariant Chern–Simons gravity, is a Carroll limit of the Liouville theory associated with three-dimensional AdS$_3$ gravity, a natural question is whether the deformation of Poincar\'e symmetry defined by the Maxwell algebra can lead to a boundary theory containing a post-Carrollian correction of such a Carrollian CFT. So far, post-Carrollian corrections to Carroll limits have been considered in the case of particle models \cite{Gomis:2022spp} and gravitational theories in various dimensions \cite{Gomis:2019nih,Hansen:2021fxi,Ecker:2025ncp}.

In this paper, we study the two-dimensional boundary theory induced at the boundary of Maxwell-invariant Chern–Simons gravity in three spacetime dimensions. After solving the constraints and imposing a suitable gauge-fixing condition, the gravitational action reduces to a Maxwell-invariant chiral Wess–Zumino–Witten (WZW) model. Imposing boundary conditions associated with the Maxwell–BMS$_3$ symmetry considered in \cite{Concha:2018zeb}, we show that this chiral WZW model further reduces to a Maxwellian extension of the flat Liouville model considered in \cite{Barnich:2012rz,Barnich:2013yka}. Finally, we consider the Carrollian expansion of the boundary theory associated with AdS$_3$ gravity \cite{Coussaert:1995zp}, including the contribution of the so-called exotic gravitational term \cite{Witten:1988hc}. We show that the boundary dual models mentioned above naturally emerge in the expansion as the terms along different powers of the speed of light, signaling a potential relation between Maxwell spacetime symmetry and post-Carrollian corrections to a Carrollian CFT.

The paper is structured as follows: Section~\ref{A} reviews the definition of the Maxwell group in three spacetime dimensions as an extended semidirect product based on the $\SL$ symmetry and introduces the basic group-theoretical tools used in the subsequent sections. Section~\ref{B} reviews Maxwell-invariant Chern–Simons gravity in three spacetime dimensions, together with the boundary conditions and asymptotic symmetry algebra given in \cite{Concha:2018zeb}, which defines a Maxwell extension of the BMS$_3$ symmetry with three central terms. In Section~\ref{C}, the Chern–Simons action is reduced to a chiral WZW model with Maxwell invariance. After imposing the boundary conditions used to obtain the asymptotic symmetry algebra, the model reduces to a Maxwellian extension of the flat Liouville theory. We show how the Maxwellian boundary dual can also be obtained from the analysis of the geometric action on coadjoint orbits of the Maxwell–BMS$_3$ group. Finally, we show how a Carrollian expansion of the known boundary dual associated with three-dimensional AdS$_3$ Chern–Simons gravity with Brown–Henneaux boundary conditions reproduces the flat Liouville model and the Maxwellian boundary dual as a post-Carrollian correction. Section~\ref{D} concludes with a discussion of our results and their potential applications and future developments.

\section{The Maxwell symmetry in three spacetime dimensions}
\label{A}
In this section, we introduce the basic definitions related to the Maxwell group and the Maxwell algebra in three spacetime dimensions that will be used in the subsequent sections. For further details on Maxwell symmetry, we refer the reader to \cite{Schrader:1972zd,Gomis:2009dm,deAzcarraga:2012qj,Salgado-Rebolledo:2019kft}.

\subsection{The Maxwell$_3$ group}
\label{A1}
The Maxwell group in 2+1 dimensions is the extended semidirect product \cite{Salgado-Rebolledo:2019kft}
\beq
{\rm Maxwell}_3= \SL\ltimes_{{\rm Ad}} \sl_{ab}^{(ext)},
\eeq
where ${\rm Ad}$ denotes the adjoint representation of the $\SL$ group, $\sl_{ab}$ is the Abelian vector space formed by the $\sl$ algebra matrices under matrix addition, and $\sl_{ab}^{(ext)}$ stands for the extension $\sl_{ab}\times \sl_{ab}$ with elements $(a,\tilde a)$, endowed with the “extended sum”
\beq\label{extsum}
(a,\tilde a)\hat{+}(b,\tilde b)=(a+b\,,\;\tilde a +\tilde b +  [a,b]).
\eeq
The elements of the Maxwell$_3$ group are therefore given by triplets $\left(\Lambda,a,\tilde a\right)$, where $\Lambda\in\SL$ and $a,\tilde a\in \sl_{ab}$. The group product is
\beq\label{productMaxwell}
(\Lambda,a,\tilde a) (\Gamma,b,\tilde b)=\left(\Lambda\Gamma\,,\; a+ \Lambda  b \Lambda^{-1} \,,\; \tilde a + \Lambda \tilde b \Lambda^{-1} + \frac12 \left[a,\Lambda b \Lambda^{-1} \right] \right).
\eeq
The associated Lie algebra, $\mathfrak{maxwell}_3$, has the structure of an extended semidirect sum, and its elements can be identified with the tangent vectors at the identity of the corresponding Lie group
\beq\label{Malgelement}
(X,a,\tilde a) =\frac{d}{ds} \left(e^{s X}, s a, s\tilde a\right)\bigg|_{s=0}.
\eeq
The Lie algebra commutation relations are obtained by evaluating the infinitesimal adjoint representation
\beq\label{infadjoint}
\frac{d}{ds_1} \frac{d}{ds_2}\bigg(
(e^{s_1X},s_1 a, s_1\tilde a)
(e^{s_2Y},s_2 b, s_2\tilde b)
(e^{s_1X},s_1 a, s_1\tilde a)^{-1}\bigg)\bigg|_{s_1=s_2=0},
\eeq
which leads to the bracket
\beq\label{maxwellbracket}
\left[(x,a,\tilde a) \;,\;(y,b,\tilde b) \right]=\left([x,y]\,,\;  [x,b]-[y,a]\,,\;[x,\tilde b]-[y,\tilde a]+[a,b]\right).
\eeq
Now we choose a basis $t_a$ for $\sl$ satisfying
\beq\label{sl2commKilling}
[t_a,t_b]=\epsilon^c{}_{ab} t_c, \qquad
{\rm Tr}(t_a t_b)=\eta_{ab},
\eeq
where ${\rm Tr}$ denotes the matrix trace, $\epsilon_{abc}$ is the Levi-Civita symbol in three dimensions, and $\eta_{ab}$ is the $\sl$ Killing form. The latter can be normalized as
\beq
\eta_{ab} = \left(
\begin{array}{c@{\quad}c@{\quad}c}
0 & 1 & 0 \\
1 & 0 & 0 \\
0 & 0 & 1
\end{array}
\right),
\eeq
and coincides with the Minkowski metric in 2+1 dimensions in light-cone coordinates. Indices will be raised and lowered with this metric and its inverse, which has the same form. Expanding the $\sl$ elements in this basis
\beq
x=x^a t_a, \qquad a=a^a t_a,\qquad \tilde a = \tilde a^a t_a, 
\eeq
and defining the generators
\beq\label{Maxwelbasis}
J_a=(t_a,0,0),\qquad
P_a=(0,t_a,0),\qquad
Z_a=(0,0,t_a),
\eeq
the elements of the $\mathfrak {maxwell}_3$ algebra can be written in the familiar form
\beq\label{expandgen}
X=(x,a,\tilde a)= x^a J_a+ a^a P_a + \tilde a^a Z_a.
\eeq
Evaluating the bracket \eqref{maxwellbracket} for the generators $J_a$,  $P_a$, and  $Z_a$ leads to the Maxwell$_3$ commutation relations \cite{Cangemi:1992ri}
\beq\label{maxwellalgebra3D}
[J_a, J_b]= \epsilon^c{}_{ab}J_c, \qquad
[J_a, P_b]= \epsilon^c{}_{ab}P_c, \qquad
[P_a, P_b]= \epsilon^c{}_{ab}Z_c= [J_a, Z_b].
\eeq
By introducing the dual generators $J_{ab}=\epsilon_{abc} J^c$ and $Z_{ab}=\epsilon_{abc} Z^c$, the commutation relations \eqref{maxwellalgebra3D} can be written in the standard form \cite{Schrader:1972zd}, which holds in four and higher dimensions. These include the non-vanishing Poincar\'e commutation relations, the standard transformation of the antisymmetric tensor $Z_{ab}$ under Lorentz transformations, and the commutation relation \eqref{MaxalgD} characteristic of the Maxwell algebra. The generalization of the construction \eqref{productMaxwell} to higher dimensions can be found in \cite{Schrader:1972zd,Salgado-Rebolledo:2019kft}.

The non-degenerate invariant bilinear form on $\mathfrak{maxwell}_3$ is given by
\beq\label{invform0}
\bal
\left\langle (x,a,\tilde a) \;,\;(y,b,\tilde b)\right\rangle
&= \kappa_1 {\rm Tr}(x,y) + \kappa_2 \Big({\rm Tr}(x,b)+{\rm Tr}(y,a)\Big)
\\&+\kappa_3 \Big({\rm Tr}(x,\tilde b)+{\rm Tr}(y,\tilde a)+{\rm Tr}(a,b)\Big).
\eal
\eeq
Using the $\sl$ Killing form \eqref{sl2commKilling}, this invariant form can be expressed in the basis \eqref{Maxwelbasis} as
\beq\label{maxwellinvform}
\left\langle J_a J_b\right\rangle = \kappa_1 \eta_{ab}, \qquad
\left\langle J_a P_b\right\rangle = \kappa_2 \eta_{ab}, \qquad
\left\langle P_a P_b\right\rangle = \kappa_3 \eta_{ab}
=\left\langle J_a Z_b\right\rangle.
\eeq
Notice that if one eliminates the third slot in the group and algebra elements featuring in Eqs.~\eqref{productMaxwell}, \eqref{maxwellbracket}, and \eqref{Maxwelbasis}, one recovers the semidirect product structure defining the Poincar\'e group and the Poincar\'e algebra in 2+1 dimensions \cite{Barnich:2014kra}. This effectively sets $Z_a=0=\kappa_3$ in the relations \eqref{maxwellalgebra3D} and \eqref{maxwellinvform}, yielding the $\mathfrak{iso}(2,1)$ commutation relations together with its associated invariant bilinear form \cite{Witten:1988hc}.

Using the product law \eqref{productMaxwell}, one can compute the Maurer–Cartan forms on the Maxwell$_3$ group. Given a group element $G(t)$ defining a curve in the group manifold with parameter $t$, and its inverse
\beq\label{Maxwellgrouplement}
G(t)=\big(\Lambda(t), a(t) , \tilde a(t)\big), \qquad G^{-1}(t)=\Big(\Lambda^{-1}(t),-\Lambda(t) a(t) \Lambda^{-1}(t),-\Lambda(t) \tilde a(t) \Lambda^{-1}(t)\Big),
\eeq
the left-invariant and right-invariant Maurer–Cartan forms are obtained by introducing another curve with parameter $\hat t$ and computing the differential of the left and right multiplication by $G^{-1}$:
\beq\label{MCformsdef}
\bal
\Xi_L&=\frac{d}{d\hat t}\Big( G^{-1}(t)G(\hat t\,)\Big)\bigg|_{\hat t=t}\d t, \\
\\
\Xi_R&=\frac{d}{d\hat t}\Big( G(\hat t\,)G^{-1}(t)\Big)\bigg|_{\hat t=t}\d t .
\eal
\eeq
Explicitly, this gives
\begin{subequations}\label{MCforms}\begin{align}
\Xi_L=G^{-1}\d G& = \left(\Lambda^{-1}\d \Lambda,\; \Lambda^{-1}\d a \Lambda ,\; \Lambda^{-1}\d\tilde a\Lambda  -\frac12\Lambda^{-1}[a,\d a]\Lambda \right),
\\
\Xi_R=\d G G^{-1}&= \left(\d \Lambda \Lambda^{-1},\;D^{-\d\Lambda\Lambda^{-1}} a   ,\; D^{-\d\Lambda\Lambda^{-1}} \tilde a+\frac12 \left[a,\;D^{-\d\Lambda\Lambda^{-1}} a\right]  \right),
\end{align}
\end{subequations}
where we have defined
\beq\label{DX}
D^x a=\d a +[x,a].
\eeq
Notice that, as a slight abuse of language, we have omitted the fact that the expressions in Eq.~\eqref{MCformsdef} represent the pullback of the Maurer–Cartan forms to the curve, while in \eqref{MCforms} we have bluntly replaced $\frac{dG}{dt}dt$ by $\d G$ in order to recover the Maurer–Cartan forms on the full group.

\subsection{Geometric action}
\label{A2}

An element that will be useful in our analysis of the boundary dynamics of Chern–Simons gravity is the geometric action associated with a coadjoint orbit of a Lie group. Here, we briefly review the main properties of such systems. For a detailed description, see \cite{Alekseev:1988ce,Aratyn:1990dj} and references therein. Applications to three-dimensional gravity can be found in Refs.~\cite{Barnich:2017jgw,Henneaux:2019sjx,Merbis:2019wgk,Merbis:2023uax} and, in particular, related to the Maxwell group, in \cite{Salgado-Rebolledo:2019kft}.

Geometric actions define dynamical systems whose symplectic structure is the Kirillov–Kostant–Souriau symplectic form on coadjoint orbits of a Lie group. In order to define them, we need to consider some Lie algebra with elements $X$, its dual space with elements denoted by $S$, and a pairing between them, namely the space of linear maps determined by the elements $S$ 
\beq\label{genpairing}
\llangle S\,,\, X \rrangle \in \mathbb R.
\eeq
Given a pairing, the adjoint action of the Lie group on its Lie algebra
\beq
{\rm Ad}_G X= \frac{d}{dt}\Big( G\,H(t)\,G^{-1}\Big)\bigg|_{t=0},\qquad X=\frac{d}{dt}H(t)\bigg|_{t=0},
\eeq
induces an action on the Lie algebra dual, called the coadjoint representation of the group and denoted by ${\rm Ad}^*_G S$, such that the pairing is invariant under the combined action
\beq\label{defadcoad}
\llangle {\rm Ad}^*_G S\,,\, {\rm Ad}_G X \rrangle= \llangle S\,,\, X \rrangle.
\eeq
Similarly, the infinitesimal version of this expression connects the adjoint action of the corresponding Lie algebra \eqref{infadjoint} with its coadjoint action
\beq
\llangle {\rm ad}^*_X S\,,\, Y \rrangle+\llangle  S\,,\, {\rm ad}_X Y \rrangle=0.
\eeq
Now we consider the case of a Maxwell-like group with elements $G=(\Lambda, a,\tilde a)$ and product law
\beq\label{GenproductMaxwell}
(\Lambda,a,\tilde a) (\Gamma,b,\tilde b)=\left(\Lambda\Gamma\,,\; a+ {\rm Ad}_\Lambda  b  \,,\; \tilde a +{\rm Ad}_\Lambda  \tilde b + \frac12 {\rm ad}_a {\rm Ad}_\Lambda b \right).
\eeq
This is an extended semidirect product based on the adjoint representation of a Lie group on its Lie algebra, and reproduces Maxwell$_3$ when that Lie group is chosen to be $\SL$. While our focus for now is the Maxwell group, this construction naturally extends to more general Maxwell-like matrix groups, for which
\begin{equation}\label{AdMatrix}
{\rm Ad}_\Lambda = \Lambda(\,)\Lambda^{-1} = {\rm Ad}^*_\Lambda, \qquad 
{\rm ad}_x = [x,\,] = {\rm ad}^*_x,
\end{equation}
and can also be generalized to infinite-dimensional extensions. Indeed, an extended semidirect product of the form \eqref{GenproductMaxwell}  based on the Virasoro algebra will be considered in Section~\ref{C2}. 

The Lie algebra associated to the extended semidirect product has elements of the form~\eqref{Malgelement}, $Y=(y,b,\tilde b)$, whereas dual Lie algebra elements will be denoted by $S=(s,p,\tilde p)$. The pairing \eqref{genpairing} is thus decomposed as
\beq
\llangle S, Y\rrangle = \llangle s, y\rrangle +\llangle p, a\rrangle+\llangle \tilde p, \tilde a\rrangle,
\eeq
where in this case $\llangle s, y\rrangle$ is the pairing between the $\sl$ algebra and its dual, and matches the Killing form \eqref{sl2commKilling}. The adjoint and the coadjoint representations then take the form
\begin{subequations}
\begin{align}
&{\rm Ad}_G Y=\bigg(
{\rm Ad}_\Lambda y
,
{\rm Ad}_\Lambda b - {\rm ad}_{{\rm Ad}_\Lambda y} a 
,\,
{\rm Ad}_\Lambda \tilde b - {\rm ad}_{{\rm Ad}_\Lambda y} \tilde a +{\rm ad}_a\Big({\rm Ad}_\Lambda b -\frac12  {\rm ad}_{{\rm Ad}_\Lambda y} a \Big)\bigg), \label{Adgen}
\\[8pt]
&{\rm Ad}^*_{G^{-1}} S=\bigg(
{\rm Ad}^*_{\Lambda^{-1}} 
\Big(s
-
{\rm ad}^*_a p 
-{\rm ad}^*_{\tilde a}\tilde p +\frac12 ({\rm ad}^*_a)^2 \tilde p\Big)
,\,
{\rm Ad}^*_{\Lambda^{-1}} ( p -{\rm ad}^*_{a}  \tilde p )
,\,
{\rm Ad}^*_{\Lambda^{-1}} \tilde p
\bigg), \label{Coadgen}
\end{align}
\end{subequations}
The coadjoint action on a fixed dual space element $S_{\rm o}$ generates a coadjoint orbit with elements $S={\rm Ad}^*_G S_{\rm o}$, given by the quotient of the Lie group and the isotropy group of $S_{\rm o}$. Such a manifold is endowed with the Kirillov–Kostant–Souriau symplectic form, which is the pullback to the orbit of the pre-symplectic structure
\beq
\Theta = \llangle S_{\rm o}\,,\, \Xi_L \wedge \Xi_L \ \rrangle = -\d \llangle S_{\rm o}\,,\, \Xi_L \ \rrangle.
\eeq
The geometric action on the orbit can be constructed out of the potential associated to $\Theta$ as
\beq\label{geometricaction1}
I_{\rm GA} = \int_{\mathcal C} \llangle S_{\rm o}\,,\, \Xi_L \ \rrangle,
\eeq
where the integration is over a path $\mathcal C$ on the orbit, and it is understood that the integrand is pulled back to the curve. Explicitly this gives
\beq\label{genactionMaxwell1}
I^{{\rm Maxwell}_3}_{\rm GA}= \int_{\mathcal C} \left( \big\langle\!\big\langle s_{\rm o},\, \Lambda^{-1}\d \Lambda \big\rangle\!\big\rangle
+\big\langle\!\big\langle {\rm Ad}^*_\Lambda p_{\rm o},\,\d a \big\rangle\!\big\rangle
+\Big\langle\!\Big\langle {\rm Ad}^*_\Lambda  \tilde p_{\rm o},\,\left(\d \tilde a -\frac12{\rm ad}_a\d a\right)
\Big\rangle\!\Big\rangle\right).
\eeq
Hence, using the left-invariant Maurer–Cartan form for the Maxwell group \eqref{MCforms} and \eqref{AdMatrix}, the geometric action takes the form
\beq\label{partactionMaxwell}
I^{{\rm Maxwell}_3}_{\rm GA}= \int_{\mathcal C} d\tau\, {\rm Tr}\left[ s_{\rm o}\, \Lambda^{-1}\frac{d \Lambda}{d\tau} + \Lambda \,p_{\rm o}\,\Lambda^{-1}\frac{d a}{d\tau}  +\Lambda \, \tilde p_{\rm o}\,\Lambda^{-1}\left(\frac{d\tilde a}{d\tau} -\frac12\Big[a,\frac{d a}{d\tau} \Big]\right) \right],
\eeq
where $\tau$ parametrizes $\mathcal C$. This action describes a spinning particle in $2+1$ dimensions \cite{deSousaGerbert:1990yp}, with coordinates $a^a$, momentum $\pi_a=(\Lambda \,p_{\rm o}\,\Lambda^{-1})_a$, spin $s_a={\rm Tr}(s_{\rm o}\, \Lambda t_a \Lambda^{-1})$, in the presence of a constant electromagnetic field with field strength $F_{ab}=\epsilon^c{}_{ab}(\Lambda \,\tilde p_{\rm o}\,\Lambda^{-1})_c$. The ``extra coordinates'' play the role of Lagrange multipliers enforcing the constraint $\frac{d}{d\tau} F_{ab}=0$ \cite{Bonanos:2008ez,Gibbons:2009me}. 

For our purposes, it will prove useful to rewrite by replacing
\beq\label{idGA}
G\rightarrow G^{-1}, \qquad
S_{\rm o}\rightarrow - S_{\rm o},
\eeq
in the definition \eqref{geometricaction1}, which in particular implies $\Xi_L\rightarrow -\Xi_R=-{\rm Ad}_G \,\Xi_L$. Using \eqref{defadcoad}, this leads to
\beq\label{geometricaction2}
I_{\rm GA} =  \int_{\mathcal C} \llangle {\rm Ad}^*_{G^{-1}}S_{\rm o}\,,\, \Xi_L \ \rrangle.
\eeq
Using \eqref{defadcoad}, and discarding total derivatives, this leads to
\beq\label{genactionMaxwell2}
I^{{\rm Maxwell}_3}_{\rm GA}=\int_{\mathcal C}
\Big\langle\!\Big\langle
{\rm Ad}^*_{\Lambda^{-1}} 
\Big(s_{\rm o}
-
{\rm ad}^*_a p_{\rm o} 
-{\rm ad}^*_{\tilde a}\tilde p_{\rm o} +\frac12 ({\rm ad}^*_a)^2 \tilde p_{\rm o}\Big)
, \Lambda^{-1} \d \Lambda
\Big\rangle\!\Big\rangle
-\frac12
\int_{\mathcal C} \Big\langle\!\Big\langle
{\rm ad}^*_a \tilde p_{\rm o}, \d a  \Big\rangle\!\Big\rangle.
\eeq
One can show, using \eqref{AdMatrix} and \eqref{idGA}, together with the explicit form of the inverse Maxwell group element \eqref{Maxwellgrouplement}, that the action \eqref{genactionMaxwell2} is equivalent to \eqref{partactionMaxwell}. However, in Section~\ref{C3}, this form of the action will prove convenient when considering infinite-dimensional generalizations. Additionally, it will be useful to consider the Noether charges associated to the global symmetries of \eqref{geometricaction2}, which are given by
\begin{equation}\label{DefQ}
Q_X = -\left\llangle {\rm Ad}^*_{G^{-1}} S_{{\rm o}},\, X \right\rrangle,
\end{equation}
with $X$ a parameter associated to an infinitesimal global transformation defined by the right action of the group $G \rightarrow G(I + X)$, and $\delta Q_X = \iota_{\mathfrak{v}_X} \Theta$, where $\mathfrak{v}_X$ is the vector field associated to the right translations that reduces to $X$ at the identity \cite{Barnich:2017jgw}. The charges satisfy a Poisson algebra isomorphic to the Lie algebra of global symmetry transformations
\begin{equation}\label{PoissonQ}
\delta_X Q_Y = \mathfrak{L}_{\mathfrak{v}_X} Q_Y = \{Q_X, Q_Y\} = Q_{{\rm ad}_X Y},
\end{equation}
where $\mathfrak{L}$ is the Lie derivative. Moreover, the geometric action is invariant under gauge transformations given by left multiplication with elements belonging to the isotropy group of the orbit representative.

\section{Maxwell$_3$-invariant  Chern-Simons gravity}
\label{B}
In this section, we review the main features of three-dimensional Chern-Simons gravity invariant under the $\mathfrak{maxwell}_3$ algebra \cite{Cangemi:1992ri}. We consider the boundary conditions introduced in Ref.~\cite{Concha:2018zeb} as an extension of the asymptotically flat construction given in \cite{Barnich:2013yka} (see also \cite{Barnich:2014cwa,Grumiller:2017sjh}). This will prepare the ground for reducing the bulk action to the boundary in the next section.

\subsection{Chern-Simons action}
\label{B1}

In three space-time dimensions, it is possible to define a Maxwell$_3$-invariant gravity action as the integral of a Chern-Simons form over some three-dimensional manifold $M$
\beq\label{CSactiongen}
I_{\rm CS}=-\int_M \left\langle
\mathcal A\wedge \d\mathcal A+ \frac23 
\mathcal A\wedge\mathcal A\wedge\mathcal A
\right\rangle,
\eeq
where $\mathcal A$ is a one-form gauge connection taking values on the Maxwell$_3$ algebra \eqref{maxwellalgebra3D}, i.e.,
\beq
\mathcal A =  \Omega^a J_a+ E^a P_a + \Sigma^a Z_a.
\eeq
The gauge fields of the theory are the spin connection $\Omega^a$, the dreibein $E^a$, and the Maxwellian field $\Sigma^a$. Adopting the notation \eqref{expandgen}, the connection can be expressed as
\beq\label{Atriad}
\mathcal A = (\Omega, E,\Sigma), \qquad
\Omega=\Omega^a t_a ,\quad E=E^a t_a,\quad\Sigma=\Sigma^a t_a.
\eeq
The corresponding curvature is computed from the definition \eqref{maxwellbracket}, which must be generalized to Lie algebra-valued $p$-forms as follows
\beq\label{bracketpforms}
\Big[
\underbrace{(X,a,\tilde a)}_{\text{$p$-form}} \;,\;
\underbrace{(Y,b,\tilde b)}_{\text{$q$-form}}
\Big]=\left([X,Y]\;,\;  [X,b]-(-1)^{pq}[Y,a]\;,\;[X,\tilde b]-(-1)^{pq}[Y,\tilde a]+[a,b]\right).
\eeq
This leads to
\beq
\mathcal F= \d \mathcal A +\frac12[\mathcal A, \mathcal A]=(R,T,F),\qquad
R=R^a t_a ,\quad T=T^a t_a,\quad F=F^a t_a,
\eeq
where the Lorentz curvature $R^a$, the torsion $T^a$, and the field strength associated with the Maxwellian field $F^a$ are respectively given by
\beq
R=d\Omega+\frac12[\Omega, \Omega], \qquad
T=dE+[\Omega, E],\qquad
F=d\Sigma+[\Omega, \Sigma]+\frac12 [E,E].
\eeq
Using \eqref{invform0}, the explicit form of the action is
\beq
\bal
I^{\rm Maxwell_3}_{\rm CS}= -\int_M \bigg[\kappa_1\, CS(\Omega)+ 2\kappa_2 \, E^a R_a + \kappa_3 \left(2\Sigma^a R_a + E^a T_a \right)\bigg],
\eal
\eeq
where $CS(\Omega)$ denotes the Lorentz Chern-Simons form constructed from the spin connection
\beq
CS(\Omega)= \Omega_a\wedge\d \Omega^a +\frac13 \epsilon_{abc}\Omega^a\wedge \Omega^b\wedge\Omega^c.
\eeq
The field equations are given by the vanishing of the curvature $\mathcal F$, i.e.
\beq\label{CSfieldeqs}
R^a =0, \qquad T^a= 0,\qquad F^a=0.
\eeq
Since the field equations governing the spacetime geometry are the same as in the $\mathfrak{iso}(2,1)$ case \cite{Barnich:2013yka}, in the next section we use the same solution space for the dreibein and spin connection.

\subsection{Boundary conditions and asymptotic symmetry algebra}
\label{B2}
We consider a space-time $M=D \times \mathbb{R}$, where $D$ plays the role of a spatial slice with disk topology and spatial coordinates $r\in[0,\infty)$, $\phi\in[0,2\pi]$, whereas $\mathbb{R}$ represents time with coordinate $u$. Following Ref.~\cite{Banados:1998gg}, we solutions of Maxwell$_3$-invariant Chern-Simons gravity \eqref{CSactiongen} of the form
\beq\label{defalpha}
\mathcal A=h^{-1}\alpha h+ h^{-1} \d h,
\eeq
where $h=h(r)$ is a Maxwell$_3$ group element that encodes the radial dependence of the connection. The boundary dynamics of the theory is encoded in a boundary connection $\alpha$, which depends only on the boundary coordinates $u$ and $\phi$. We define it as \cite{Concha:2018zeb}
\beq\label{boundaryconnection}
\alpha=
 \frac12 \mathcal M \d\phi J_0+\d\phi J_1
 + \frac12\left(\mathcal M \d u+ \mathcal N \d\phi\right) P_0  +\d u P_1 
 +\frac12\left(\mathcal N \d u + \mathcal R d\phi\right) Z_0,
\eeq
with
\beq\label{functionsMaxwell}
\mathcal M=\mathcal M(\phi),\qquad
\mathcal N= \mathcal T(\phi) + u \mathcal M^\p (\phi),\qquad
\mathcal R= \mathcal Y(\phi) + u \mathcal T^\p (\phi)+ \frac{u^2}{2} \mathcal M^{\p\p}(\phi),
\eeq
where the prime denotes the derivative with respect to $\phi$, and $\mathcal M$, $\mathcal T$ and $\mathcal Y$ are arbitrary functions of $\phi$. To maintain generality, we do not fix the form of $h$, but an important example is $h = e^{r P_0}$, for which the dreibein and the spin connection in \eqref{defalpha} are given by asymptotically flat solutions of the Einstein equations in the BMS gauge \cite{Barnich:2013yka}, whereas the Maxwellian field $\Sigma$ becomes a solution of the equation $F^a=0$ in \eqref{CSfieldeqs} on those geometries \cite{Concha:2018zeb}. The gauge transformations $\delta \alpha=\d \varepsilon+[\alpha,\varepsilon]$, with $\varepsilon=\varepsilon^a_J J_a+\varepsilon^a_P P_a + \varepsilon^a_Z Z_a$, that preserve the form of $a$, can be expressed in terms of three functions $Y_i=Y_i(\phi)$, $i=1,2,3$, as
\begin{subequations}\label{solparameters}
\begin{align}
\varepsilon_J^0&= \frac{\mathcal M}2 \varepsilon_J^1-(\varepsilon_J^1)^{\p\p},
&
\varepsilon_J^1&= Y_1,
& 
\varepsilon_J^2&=-(\varepsilon_J^1)^\p,
\\
\varepsilon_P^0&= \frac{\mathcal M}2 \varepsilon_P^1 + \frac{\mathcal N}2 \varepsilon_J^1-(\varepsilon_P^1)^{\p\p},
&
\varepsilon_P^1&= Y_2+uY_1^\p,
& 
\varepsilon_P^2&=-(\varepsilon_P^1)^\p,
\\
\varepsilon_Z^0&= \frac{\mathcal M}2 \varepsilon_Z^1 + \frac{\mathcal N}2 \varepsilon_P^1+\frac{\mathcal R}2 \varepsilon_J^1-(\varepsilon_Z^1)^{\p\p},
&
\varepsilon_Z^1&= Y_3+uY_2^\p+\frac{u^2}2Y_1^{\p\p},
& 
\varepsilon_Z^2&=-(\varepsilon_Z^1)^\p,
\end{align}
\end{subequations}
Comparing the variation of $ \alpha$ defined by these gauge transformations with the variation of \eqref{boundaryconnection}, one finds that the angular-dependent functions $\mathcal M$, $\mathcal T$, and $\mathcal Y$ transform as
\begin{subequations}\label{deltaMTY}
\begin{align}
&\delta \mathcal M= \mathcal M^\p Y_1 + 2 \mathcal M Y_1^\p - 2 Y_1^{\p\p\p},
\\
&\delta \mathcal T= \mathcal T^\p Y_1 + 2\mathcal T Y_1^\p + \mathcal M^\p Y_2 + 2 \mathcal M Y_2^\p - 2 Y_2^{\p\p\p} ,
\\
&\delta \mathcal Y= \mathcal Y^\p Y_1 + 2\mathcal Y Y_1^\p +  \mathcal T^\p Y_2 + 2\mathcal T Y_2^\p + \mathcal M^\p Y_3 + 2 \mathcal M Y_3^\p - 2 Y_3^{\p\p\p} .
\end{align}
\end{subequations}
For an asymptotic gauge variation, the variation of the Chern-Simons charge obtained through the Regge-Teitelboim prescription \cite{Regge:1974zd} is given by \cite{Banados:1994tn}
\begin{equation}
\delta Q[\varepsilon] = 2 \int d\phi \, \left\langle \varepsilon \, \delta \alpha_\phi \right\rangle.
\end{equation}
After substituting \eqref{boundaryconnection} and \eqref{solparameters}, this expression can be integrated, yielding three independent sets of charges 
\begin{subequations}
\begin{align}
j[Y_1] &= \int d\phi \, \left( \kappa_3 \mathcal{Y} + \kappa_2 \mathcal{T} + \kappa_1 \mathcal{M} \right) Y_1, \\
p[Y_2] &= \int d\phi \, \left( \kappa_3 \mathcal{T} + \kappa_2 \mathcal{M} \right) Y_2, \\
z[Y_3] &= \int d\phi \, \kappa_3 \mathcal{M} \, Y_3.
\end{align}
\end{subequations}
Expanding the charges in Fourier modes as
\beq
j_m = j[e^{im\phi}], \quad p_m = p[e^{im\phi}], \quad z_m = z[e^{im\phi}],
\eeq
and computing their Dirac brackets using the relation
\begin{equation}\label{brackelcharges}
\{ Q[\varepsilon_1], Q[\varepsilon_2] \} = \delta_{\varepsilon_1} Q[\varepsilon_2],
\end{equation}
one finds the Maxwell extension of the (centrally extended) $\mathfrak{bms}_3$ algebra \cite{Concha:2018zeb} 
\begin{subequations}\label{maxwellbms3alg}
\begin{align}
i \{ j_m, j_n \} &= (m - n) j_{m+n} + \frac{c_1}{12} m^3 \delta_{m+n,0}, \\
i \{ j_m, p_n \} &= (m - n) p_{m+n} + \frac{c_2}{12} m^3 \delta_{m+n,0}, \\
i \{ j_m, z_n \} &= (m - n) z_{m+n} + \frac{c_3}{12} m^3 \delta_{m+n,0}, \\
i \{ p_m, p_n \} &= (m - n) z_{m+n} + \frac{c_3}{12} m^3 \delta_{m+n,0},
\end{align}
\end{subequations}
also referred to as the $\mbms$ algebra for short, where
\beq \label{candkappa}
c_i = 48 \pi \kappa_i, \quad i=1,2,3.
\eeq
This algebra was previously derived in a purely algebraic framework as an expansion of the Virasoro algebra \cite{Caroca:2017onr}.


\section{Boundary dynamics}
\label{C}
In order to write the Chern-Simons action in Hamiltonian form, temporal and spatial indices are split as
\beq
\mathcal A= \mathcal  A_\mu dx^\mu= \mathcal  A_u \d u +\mathcal  A_r \d r +\mathcal  A_\phi \d \phi ,
\eeq
Considering $\epsilon^{ur\phi}\equiv -1$, the Chern-Simons action \eqref{CSactiongen} can be written as
\begin{equation}
I_{\rm CS}=\int d^3x \;\left\langle 2\mathcal A_u \mathcal F_{r\phi}- \mathcal A_r \dot{\mathcal A}_\phi + \mathcal A_\phi \dot{\mathcal A}_r\right\rangle  
+B,
\end{equation}
where the dot denotes the derivative with respect to $u$, and $B$ is some boundary term such that the variation of $I$ is well-defined. Varying the action with respect to the spatial components of the connection gives
\begin{equation}
\delta I=2\int d^3 x 
\left\langle
\mathcal F_{ur} \delta \mathcal A_\phi -  \mathcal F_{u\phi} \delta \mathcal A_r
\right\rangle
+2\int dud\phi \left\langle \mathcal A_u \delta \mathcal A_\phi \right\rangle
 + \delta B,
\end{equation}
where we have defined the surface element  as $ds_\mu = du d\phi \, n_\mu$, with $n_\mu=(0,1,0)$ the unit vector normal to the boundary. Therefore, the condition on $B$ is
\beq\label{varBterm}
\delta B =-2\int d u d \phi \left\langle \mathcal A_u\delta \mathcal A_\phi\right\rangle,
\eeq
On the other hand, variation of the action with respect to $\mathcal A_u$ leads to the constraint $\mathcal F_{ij}=0$, whose local solution is
\beq
\mathcal A_i=   G^{-1} \partial_i G\,,
\eeq
with $G$ an element belonging to the gauge group. We also adopt the partial gauge condition considered in Ref.~\cite{Banados:1994tn}
\begin{equation}
\partial_\phi \mathcal A_r =0\quad \Rightarrow\quad
G(u,r,\phi)= g(u,\phi) h(u,r)\,, \label{gsplitintophiandr}
\end{equation}
which leads to
\beq\label{ArAphi}
\mathcal A_r = h^{-1} \partial_r h,\qquad \mathcal A_\phi= h^{-1} \;g^{-1} g^\p\; h\,.
\eeq
Replacing these expressions back into the action together with the condition that $\dot h=0$ asymptotically, leads to the chiral WZW action \cite{Coussaert:1995zp}
\begin{equation}\label{wzwaction}
I_{\rm WZW}= \int du d\phi \, \left\langle g^{-1}\dot g \,g^{-1} g^\p \right\rangle +\frac{1}{3}\int_M\left\langle\left(G^{-1}d G\right)^3\right\rangle
+ B.
\end{equation}
Moreover, in many explicit examples considering a set of solutions in a particular gauge \cite{Banados:1998gg,Barnich:2013yka,Grumiller:2016pqb,Grumiller:2017sjh}, including the solutions in the BMS gauge defined by \eqref{defalpha}, the condition $\dot h=0$ holds not only asymptotically as in \cite{Coussaert:1995zp}, but everywhere, which implies
\beq
\mathcal A_u=h^{-1}\alpha_u h.
\eeq
Putting this together with \eqref{ArAphi} and \eqref{defalpha} one finds
\beq\label{alpha1}
\alpha = \alpha_u \d u + g^{-1} g^\p \d\phi.
\eeq
Thus, the variation \eqref{varBterm} can be put in the form
\begin{equation}\label{deltaB}
\delta B =-2\int du d\phi \left\langle \alpha_u\, \delta (g^{-1} g^\p)\right\rangle.
\end{equation}

Now, let us consider elements of the Maxwell$_3$ group as defined in Section~\ref{A1}. The elements $ g(u,\varphi) $ and $ h(r) $ can be expressed as triplets
\beq\label{gandh}
g(u,\varphi) = \big(\lambda(u,\varphi), \beta(u,\varphi), \gamma(u,\varphi)\big), \qquad
h(r) = \big(h_1(r), h_2(r), h_3(r)\big),
\eeq
or equivalently, for $ G = (\Lambda, a, \tilde{a}) $, as
\beq\label{tripletsGatildea}
\Lambda = \lambda(u,\varphi) \, h_1(r), \qquad 
a = \beta(u,\varphi) \, h_2(r), \qquad
\tilde{a} = \gamma(u,\varphi) \, h_3(r),
\eeq
where $\lambda$ and $h_1$ belong to the $SL(2,\mathbb{R})$ group, while $\beta$, $\gamma$, $h_2$, and $h_3$ are elements of the $\mathfrak{sl}(2,\mathbb{R})$. Expressing the boundary connection $\alpha$ as a triad analogously to \eqref{Atriad},
\beq
\alpha = (e, \omega, \sigma), \qquad
\omega = \omega^a t_a, \quad e = e^a t_a, \quad \sigma = \sigma^a t_a,
\eeq
the boundary conditions \eqref{boundaryconnection} can be written as
\beq\label{Bcseomegasigma}
e_u = \omega_\phi, \qquad \omega_u = 0, \qquad \sigma_u = e_\phi.
\eeq
Using the definition of the left-invariant Maurer–Cartan form given in Eq.~\eqref{MCforms}, and the decomposition in \eqref{gandh}, one finds for the $r$-independent group element $g$
\beq\label{leftMCg}
g^{-1} \d g = \big(\lambda^{-1} \d \lambda, \;\lambda^{-1} \d \beta \, \lambda, \;\lambda^{-1} \d \gamma \, \lambda - \tfrac{1}{2} \lambda^{-1} [\beta, \d \beta] \lambda \big),
\eeq
where in these expressions, $\d$ is naturally restricted to the boundary differentials $\d u$ and $\d \phi$. Hence, using Eq.~\eqref{Bcseomegasigma}, the components of $\alpha$ can be written as
\beq\label{alpha1}
\begin{aligned}
\alpha_u &= \big(0, \lambda^{-1} \lambda^\p, \, \lambda^{-1} \beta^\p \lambda \big), \\
\alpha_\phi &= g^{-1} g^\p = \big(\lambda^{-1} \lambda^\p, \, \lambda^{-1} \beta^\p \lambda, \, \lambda^{-1} \gamma^\p \lambda - \tfrac{1}{2} \lambda^{-1} [\beta, \beta^\p] \lambda \big).
\end{aligned}
\eeq
Replacing the bilinear form \eqref{invform0} in Eq. \eqref{deltaB}, the variation of $B$ can be solved to find
\begin{equation}
B = - \int du \, d\phi \, {\rm Tr} \bigg[ \kappa_2 \, \lambda^{-1} \lambda^\p \, \lambda^{-1} \lambda^\p - 2 \kappa_3 \, \lambda^\p \lambda^{-1} \beta^\p \bigg].
\end{equation}
Similarly, using \eqref{leftMCg} and the analog expression for $ G^{-1} \d G $ together with \eqref{tripletsGatildea}, we arrive at the following form of the Maxwell-WZW model action \eqref{wzwaction} 
\beq\label{MaxwellWZWaction}
\begin{aligned}
I^{\rm Maxwell_3}_{\rm WZW} &= \int du \, d\phi \,{\rm Tr}\bigg[
\kappa_1 \, \lambda^{-1} \dot{\lambda} \, \lambda^{-1} \lambda^\p + \kappa_2 \left( 2 \dot{\lambda} \lambda^{-1} \beta^\p - \lambda^{-1} \lambda^\p \, \lambda^{-1} \lambda^\p \right) \\
& \quad + \kappa_3 \left( 2 \dot{\lambda} \lambda^{-1} \gamma^\p + \beta^\p D_u^{-\dot{\lambda} \lambda^{-1}} \beta - 2 \lambda^\p \lambda^{-1} \beta^\p \right)
\bigg]
+ \frac{\kappa_1}{3} \int {\rm Tr}\left[(\Lambda^{-1} \d \Lambda)^3 \right],
\end{aligned}
\eeq
where, for simplicity and in analogy with \eqref{DX}, we have introduced the notation \cite{Barnich:2013yka}
\beq
D_\alpha^{-\partial_\alpha \lambda \lambda^{-1}} \beta \equiv \partial_\alpha \beta - [\partial_\alpha \lambda \lambda^{-1}, \beta],
\eeq
with $ x^\alpha = \{u, \phi\} $ denoting the boundary coordinates. This chiral WZW model is invariant under the loop extension of the Maxwell algebra. Indeed, the action is invariant under the following global symmetry transformations with angle-dependent parameters $ \Theta_k = \Theta_k(\phi) $, $ k=1,2,3 $.
\begin{subequations}
\begin{align}
&\lambda\smallarrow\lambda\Theta_1^{-1},
&&\beta\smallarrow\beta-u\lambda\Theta_1^{-1}\Theta_1^\p\lambda^{-1},
&& \gamma\smallarrow \gamma -\left(\frac{u}{2}{\rm ad}_\beta+\frac{u^2}{2}D_\phi^{-\lambda^\p\lambda^{-1}}\right)\lambda\Theta^{-1}_1\Theta_1^\p\lambda^{-1}
\\[6pt]
&\lambda\smallarrow\lambda,
&&\beta\smallarrow\beta+\lambda\Theta_2\lambda^{-1},
&&\gamma\smallarrow\gamma+\frac{1}{2}\left[\beta,\lambda\Theta_2\lambda^{-1}\right]
\\[6pt]
&\lambda\smallarrow\lambda,
&&\beta\smallarrow\beta,
&&\gamma\smallarrow\gamma+\lambda\Theta_3\lambda^{-1}
\end{align}
\end{subequations}
For infinitesimal transformations $ \Theta_i \sim I + \theta_i $, these become
\begin{subequations}\label{infglobal}
\begin{align}
\delta_{\theta_1} \lambda &= -\lambda \theta_1, &
\delta_{\theta_1} \beta &= - u \, \lambda \theta_1^\p \lambda^{-1}, &
\delta_{\theta_1} \gamma &= - \left( \frac{u}{2} {\rm ad}_\beta + \frac{u^2}{2} D_\phi^{-\lambda^\p \lambda^{-1}} \right) \lambda \theta_1^\p \lambda^{-1}\\
\delta_{\theta_2} \lambda &= 0, &
\delta_{\theta_2} \beta &= \lambda \theta_2 \lambda^{-1}, &
\delta_{\theta_2} \gamma &= \frac{1}{2} [\beta, \lambda \theta_2 \lambda^{-1}]\\[6pt]
\delta_{\theta_3} \lambda &= 0, &
\delta_{\theta_3} \beta &= 0, &
\delta_{\theta_3} \gamma &= \lambda \theta_3 \lambda^{-1}
\end{align}
\end{subequations}
and the associated Noether currents are obtained using the standard formula
\beq
j^\alpha[\theta_i] = -k_i^\alpha + \frac{\partial \mathcal{L}}{\partial(\partial_\alpha \psi^j)} \delta_{\theta_i} \psi^j, \qquad \text{with} \quad \delta_{\theta_i} \mathcal{L} = \partial_\alpha k_i^\alpha,
\eeq
where $\mathcal{L}$ is the Lagrangian density such that
$
I^{\rm Maxwell}_{\rm WZW} = \int du \, d\phi \, \mathcal{L},
$
and $\psi^j = \{\lambda, \beta, \gamma\}$ collectively denotes the boundary fields entering \eqref{MaxwellWZWaction}. This leads to the following currents
\beq
j^\alpha[\theta_1]=\delta^\alpha_u \,{\rm Tr}(J\theta_1),\qquad
j^\alpha[\theta_2]=\delta^\alpha_u \,{\rm Tr}(P\theta_2),\qquad
j^\alpha[\theta_3]=\delta^\alpha_u \,{\rm Tr}(Z\theta_3),
\eeq
where we have defined
\begin{subequations}\label{ChargesMKM}
\begin{align}
J&=2\kappa_1\,\lambda^{-1}\lambda^\p
+2\kappa_2\,\lambda^{-1}\beta^\p \lambda
+2\kappa_3\, \lambda^{-1}\left(\gamma^\p-\frac{1}{2}[\beta,\beta^\p ]\right)\lambda
\\
&-2\kappa_2\,u (\lambda^{-1}\lambda^\p )^\p
-2\kappa_3\, u\lambda^{-1} D^{-\lambda^\p \lambda^{-1}}_\phi\beta^\p \lambda+\kappa_3\, u^2(\lambda^{-1}\lambda^\p )^{\p\p},\nonumber
\\[6pt]
P&= 2\kappa_2\,\lambda^{-1}\lambda^\p  +2\kappa_3\,\lambda^{-1}\beta^\p \lambda -2\kappa_3\, u(\lambda^{-1}\lambda^\p )^\p,
\\[6pt]
Z&=2\kappa_3\,\lambda^{-1}\lambda^\p .
\end{align}
\end{subequations}
By introducing the Noether charges
\begin{subequations}
\begin{align}
\mathcal J[\theta_1]&=\int d\phi\, j^u[\theta_1]=\int d\phi\, {\rm Tr}(J\theta_1),
\\
\mathcal P[\theta_2]&=\int d\phi\, j^u[\theta_2]=\int d\phi\, {\rm Tr}(P\theta_2),
\\
\mathcal Z[\theta_3]&=\int d\phi\, j^u[\theta_3]=\int d\phi\, {\rm Tr}(Z\theta_3),
\end{align}
\end{subequations}
and using the Dirac bracket formula \eqref{brackelcharges} together with the infinitesimal transformations \eqref{infglobal}, the following algebra can be read off
\beq\label{MaxwellKM}
\bal
\big\{\mathcal J[\theta_1], \mathcal J[\tilde \theta_1]\big\}=&\mathcal J\big[[\theta_1,\tilde \theta_1]\big]+2\kappa_1\int d\phi \,{\rm Tr}\big(\theta_1^\p \tilde \theta_1\big),
\\
\big\{\mathcal J[\theta_1], \mathcal P[\theta_2]\big\}=&\mathcal P\big[[\theta_1, \theta_2]\big]+2\kappa_2\int d\phi \,{\rm Tr}\big(\theta_1^\p \theta_2\big),
\\
\big\{\mathcal J[\theta_1], \mathcal Z[\theta_3]\big\}=&\mathcal Z\big[[\theta_1, \theta_3]\big]+2\kappa_3\int d\phi \,{\rm Tr}\big(\theta_1^\p \theta_3\big),
\\
\big\{\mathcal P[\theta_2], \mathcal P[\tilde \theta_2]\big\}=&\mathcal Z\big[[\theta_2,\tilde \theta_2]\big]+2\kappa_3\int d\phi \,{\rm Tr}\big(\theta_2^\p \tilde \theta_2\big),
\eal\eeq
which corresponds to the Maxwell-Kac-Moody algebra \cite{Caroca:2017onr}. Notice that the action \eqref{MaxwellWZWaction} is furthermore invariant under gauge transformations with parameter $\epsilon=\epsilon(u)$ of the form $\lambda\rightarrow \epsilon\lambda$, $\beta\rightarrow \epsilon\beta\,\epsilon^{-1}$, $
\gamma\rightarrow \epsilon\gamma\,\epsilon^{-1}$. 

We now look at the field equations. Varying the action with respect to the fields $\gamma$, $\beta$, and $\lambda$ yields
\begin{subequations}
\begin{align}
&\left(\dot \lambda \lambda^{-1}\right)^\p=0,
\label{eqlambda}
\\
&D_u^{-\dot\lambda \lambda^{-1}}\beta^\p=
\left(\lambda^\p \lambda^{-1}\right)^\p,
\label{eqalpha}
\\
&D_u^{-\dot\lambda \lambda^{-1}}\left(\gamma^\p -\frac12 [\beta,\beta^\p]\right) = D_\phi^{-\lambda^\p \lambda^{-1}} \beta^\p.
\label{eqgamma}
\end{align}
\end{subequations}
The solution to Eq.~\eqref{eqlambda} reads
\beq\label{sollambda}
\lambda = \mu (u) \,\nu (\phi),
\eeq
where $\mu$ and $\nu$ are $\SL$ elements that depend arbitrarily on their respective arguments. Replacing this expression in Eq.~\eqref{eqalpha} leads to the following solution for $\beta$
\beq\label{solalpha}
\beta= \mu (u) \left( f_1 (\phi) + f_2 (u) + u \,\nu^\p(\phi) \nu^{-1}(\phi) \right) \mu^{-1}(u),
\eeq
where $f_1=f_1(\phi)$ and $f_2=f_2(u)$ are arbitrary elements of $\sl$. Finally, plugging \eqref{sollambda} and \eqref{solalpha} into \eqref{eqgamma}, we find that the solution for the field $\gamma$ reads
\beq
\bal
\gamma&= \mu(u)\bigg( 
f_3(\phi) +f_4(u) + \frac12 [f_2(u),f_1(\phi)]
+\frac{u}{2} f_1^\p(\phi)
\\
&
+ \frac{u}{2} [f_1(\phi)+f_2(u), \nu^\p(\phi)\nu^{-1}(\phi)]
+\frac{u^2}{2} \left(\nu^\p(\phi) \nu^{-1}(\phi)\right)^\p
\bigg) \mu^{-1} (u),
\eal
\eeq
where, as before, $f_3=f_3(\phi)$ and $f_4=f_4(u)$ are elements of $\sl$.

\subsection{Maxwellian boundary dual theory}
\label{C1}
To complete the reduction of the Maxwell$_3$-invariant Chern-Simons action to the boundary, we further reduce the WZW model \eqref{MaxwellWZWaction} by imposing the boundary conditions \eqref{boundaryconnection}. When expressing the boundary connection $\alpha$ as in \eqref{alpha1}, the boundary conditions become
\begin{subequations}
\begin{align}
&\lambda^{-1} \lambda^\p = \frac{\mathcal{M}}{2} t_0 + t_1,
\label{Bconds1}
\\
&\lambda^{-1} \beta^\p \lambda = \frac{\mathcal{N}}{2} t_0 ,
\label{Bconds2}
\\
&\lambda^{-1}\left( \gamma^\p - \frac{1}{2}[\beta, \beta^\p] \right)\lambda = \frac{\mathcal{R}}{2} t_0 .
\label{Bconds3}
\end{align}
\end{subequations}
Now we consider the following parametrization for the $\SL$ elements 
\beq
 \lambda = e^{-\rho_{0} t_{1}} e^{-\rho_{2} t_{2}} e^{-\rho_{1} t_{0}}.
\eeq
In terms of the fields $\rho_a$, the components of the $\SL$ Maurer–Cartan forms for $\lambda$ read
\begin{subequations}
\begin{align}
\lambda^{-1} \d\lambda &= -\left(\d\rho_{1} - \rho_{1} \d\rho_{2} - \frac{1}{2} e^{-\rho_{2}} \rho_{1}^{2} \d\rho_{0}\right) t_{0}
- e^{-\rho_{2}} \d\rho_{0} t_{1}
- \left(\d\rho_{2} + e^{-\rho_{2}} \rho_{1} \d\rho_{0}\right) t_{2},
\label{linvSL}
\\
\d\lambda \lambda^{-1} &=
- e^{-\rho_{2}} \d\rho_{1} t_{0}
+ \left(-\d\rho_{0} + \rho_{0} \d\rho_{2} + \frac{1}{2} e^{-\rho_{2}} \rho_{0}^{2} \d\rho_{1}\right) t_{1}
- \left(\d\rho_{2} + e^{-\rho_{2}} \rho_{0} \d\rho_{1}\right) t_{2}.
\label{rinvSL}
\end{align}
\end{subequations}
Using \eqref{linvSL}, restricted to $\d \rightarrow \partial_\phi \d\phi$,
the first set of boundary conditions \eqref{Bconds1} imply
\beq\label{Bconds1rho}
e^{-\rho_2} \rho_0^\p = -1,
\qquad
\rho_2^\p = \rho_1,
\qquad
- \rho_1^\p + \frac{1}{2} \rho_1^{2} = \frac{\mathcal{M}}{2}.
\eeq
On the other hand, the adjoint action of $\SL$ on $\sl$ vectors $z = z^{a} t_{a} = \eta^{ab} z_{a} t_{b}$ takes the form
\beq\label{adjointSL}
\begin{aligned}
&\lambda^{-1} z \lambda = \left(e^{\rho_{2}} z_{1} - \rho_{1} z_{2} + \rho_{1} \rho_{0} z_{1} - \frac{1}{2} \rho_{1}^{2} e^{-\rho_{2}} \left(z_{0} + \rho_{0} z_{2} - \frac{1}{2} \rho_{0}^{2} z_{1}\right)\right) t_{0}
\\
& + e^{-\rho_{2}} \left(z_{0} + \rho_{0} z_{2} - \frac{1}{2} \rho_{0}^{2} z_{1}\right) t_{1}
+ \left(z_{2} - \rho_{0} z_{1} + \rho_{1} e^{-\rho_{2}} \left(z_{0} + \rho_{0} z_{2} - \frac{1}{2} \rho_{0}^{2} z_{1}\right)\right) t_{2}.
\end{aligned}
\eeq
Using \eqref{adjointSL} for $z = \beta^\p$ allows us to express \eqref{Bconds2} as the following second set of boundary conditions 
\begin{subequations}
\begin{align}
\beta_{0}^\p + \rho_{0} \beta_{2}^\p - \frac{1}{2} \rho_{0}^{2} \beta_{1}^\p &= 0,
\\
\beta_{2}^\p - \rho_{0} \beta_{1}^\p &= 0,
\\
\left(e^{\rho_{2}} + \rho_{1} \rho_{0}\right) \beta_{1}^\p - \rho_{1} \beta_{2}^\p& = \frac{\mathcal{N}}{2},
\end{align}
\end{subequations}
which can be simplified to
\beq\label{betaprimes}
\beta_{2}^\p = \rho_{0} \beta_{1}^\p,
\qquad
\beta_{0}^\p = -\frac{1}{2} \rho_{0}^{2} \beta_{1}^\p,
\qquad
\beta_{1}^\p = e^{-\rho_{2}} \frac{\mathcal{N}}{2}.
\eeq
Similarly, using \eqref{adjointSL} for $z = \gamma^\p - \frac{1}{2} [\beta, \beta^\p]$ and the relations \eqref{betaprimes} leads to a third set of boundary conditions coming from \eqref{Bconds3} 
\begin{subequations}\label{gammaprimes}
\begin{align}
&\gamma_{0}^\p + \rho_{0} \gamma_{2}^\p - \frac{1}{2} \rho_{0}^{2} \gamma_{1}^\p = 0,
\\
&\gamma_{2}^\p - \rho_{0} \gamma_{1}^\p = - \frac{1}{2} \left(\beta_{0} + \rho_{0} \beta_{2} - \frac{1}{2} \rho_{0}^{2} \beta_{1}\right) \beta_{1}^\p,
\\
&\left(e^{\rho_{2}} + \rho_{1} \rho_{0}\right) \gamma_{1}^\p - \rho_{1} \gamma_{2}^\p = \frac{\mathcal{R}}{2} + \frac{\mathcal{N}}{4} (\beta_{2} - \rho_{0} \beta_{1}) + \frac{1}{2} \rho_{1} \left(\beta_{0} + \rho_{0} \beta_{2} - \frac{1}{2} \rho_{0}^{2} \beta_{1}\right) \beta_{1}^\p.
\end{align}
\end{subequations}
Plugging in \eqref{Bconds1rho}, \eqref{betaprimes} and \eqref{gammaprimes} into the Maxwell–WZW model \eqref{MaxwellWZWaction} and introducing the following redefinitions for the fields
\beq
\varphi = \rho_2, \qquad \xi = 2(\rho_0 \beta_1 - \beta_2), \qquad \chi = 2(\rho_0 \gamma_1 - \gamma_2) - \left(\beta_0 + \rho_0 \beta_2 - \frac{1}{2} \rho_0^2 \beta_1 \right) \beta_1,
\eeq
the action reduces to the following Maxwellian boundary dual 
\beq\label{maxwellboundarydual}
I_{\rm MBD} = \int du d\phi \, \left(
\kappa_1 \varphi^\p \dot{\varphi}
+ \kappa_2 (\xi^\p \dot{\varphi} - \varphi^\p \varphi^\p)
+ \kappa_3 \Big( \chi^\p \dot{\varphi} + \frac{1}{4} \dot{\xi} \xi^\p - \xi^\p \varphi^\p \Big)
\right).
\eeq
The term proportional to $\kappa_2$ in this action is the flat Liouville model introduced by Gomberoff, Barnich, and Gonz\'alez in Ref.~\cite{Barnich:2012rz}, whereas the extension including $\kappa_1$ was described in \cite{Barnich:2015sca, Barnich:2017jgw}. The $\kappa_2$ extension arises completely from the Maxwell extension of the Poincar\'e algebra. In the next section, we show that this action is invariant under the Maxwell-BMS$_3$ group and its conserved charges satisfy the Poisson algebra \eqref{maxwellbms3alg}.

\subsection{Maxwell-BMS$_3$ geometric action}
\label{C2}

The Lie group whose Lie algebra bracket is isomorphic to the Poisson algebra \eqref{maxwellbms3alg} is given by the extended semidirect product $\Vir\ltimes \vir_{ab}^{(ext)}$, where $\Vir$ is the Virasoro group and $\vir_{ab}$ is the Abelian vector space formed by the elements of the Virasoro algebra. Such a structure has been identified in Ref.~\cite{Salgado-Rebolledo:2019kft} as the (centrally extended) Maxwell-BMS$_3$ group. In this section, we show that the Maxwellian boundary dual model \eqref{maxwellboundarydual} is a geometric action on coadjoint orbits of the Maxwell-BMS$_3$ group and, therefore, is automatically invariant under global Maxwell-BMS$_3$ transformations.

The Virasoro group is the centrally extended diffeomorphism group of the circle, $Diff(S^1)$ \cite{Witten:1987ty}. The Lie algebra associated with $\Vir$ is the centrally extended algebra of vector fields on $S^1$, denoted $\vir$. Starting from the extended semidirect product construction \eqref{GenproductMaxwell}, it is possible to determine the Maxwell-BMS$_3$ group product law by replacing Lie group and Lie algebra elements with centrally extended pairs
\beq\label{pairs1}
\Lambda \rightarrow (U, m_1), \qquad x \rightarrow (x, m_1), \qquad
 \qquad a\rightarrow (v, m_2), \qquad \tilde a\rightarrow (\tilde v, m_3),
\eeq
where now $U$ is a diffeomorphism of the circle, $\phi\rightarrow U(\phi)$, whereas $x$, $v$ and $\tilde v$ are vector fields on $S^1$ of the form $v=v(\phi)\partial_\phi$, and $m_i$, $i=1,2,3$, are real numbers. The group product \eqref{GenproductMaxwell} turns into
\beq
(U,m_1,v,m_2,\tilde v,m_3)\bullet(V,n_1,w,n_2,\tilde w,n_3)
=(W,q_1,f,q_2,\tilde f,q_3),
\eeq
with components
\begin{subequations}\label{structureCEM}
\begin{align}
(W,q_1)&=(U,m_1)(V,n_1)
=\Big(U V,\, m_1 + n_1 +\mathcal{K} [U,V] \Big),
\\
(f,q_2)&=(v,m_2)+ {\rm Ad}_{(U,m_1)} (w,n_2) ,
\\
(\tilde f,q_3)&=(\tilde v,m_3) + {\rm Ad}_{(U,m_1)}(\tilde w,n_3)  
+ \tfrac{1}{2} {\rm ad}_{(v,m_2)} {\rm Ad}_{(U,m_1)} (w,n_2) ,
\end{align}
\end{subequations}
where $UV\equiv U\circ V$, with $\circ$ denoting function composition, and $\mathcal K [U,V]$ is the Thurston–Bott cocycle 
\beq
\mathcal K [U,V]=-\frac1{48\pi}\int_0^{2\pi} d\phi\; \log\left[U^\p\circ V\right]\left(\log\left[V^\p\right]\right)^\p.
\eeq
Here, ${\rm Ad}_{(U,m_2)}$ and ${\rm ad}_{(v,m_2)}$ denote the adjoint representations of the Virasoro group and the Virasoro algebra, respectively \cite{Witten:1987ty} 
\begin{subequations}\label{AdVir}
\begin{align}
{\rm Ad}_{(U,m_1)} (v,n_1)&= \Big({\rm Ad}_{U}v,\, n_1 -\llangle \mathcal S(U), v\rrangle \Big),
\\
{\rm ad}_{(x,m_1)} (y,n_1)&= \Big({\rm ad}_x y, \, -\llangle\mathfrak s(x),y\rrangle \Big),
\end{align}
\end{subequations}
where ${\rm Ad}_{U}v$ and ${\rm ad}_x y$ are the adjoint representations of $Diff(S^1)$ and its Lie algebra. $\mathcal S(U)$ is the Souriau cocycle, given by the Schwarzian derivative, and $\mathfrak s(x)$ is its differential at the identity. The pairing $\llangle\,,\,\rrangle$ between $\vir$ and its dual space (with elements $p=p(\phi)\d\phi^2$) is defined by integration over $S^1$ 
\begin{subequations}
\begin{align}
&{\rm Ad}_{U}v= \frac{v\circ U^{-1}(\phi)}{(U^{-1})^\p(\phi)}\partial_\phi
,\qquad\qquad {\rm ad}_x y=\left(-x(\phi) y^\p(\phi)+y(\phi) x^\p(\phi)\right)\partial_\phi ,\label{AdDiffS1}
\\
&\mathcal S(U) =\frac{1}{24\pi}\left(\frac{U^{^{\p \p \p }}}{U^{^{\p }}}-\frac{3%
}{2}\left( \frac{U^{^{\p \p }}}{U^{^{\p }}}\right) ^{2}\right), \qquad \mathfrak s(v)=\frac{1}{24\pi}v^{\p\p\p},
\\
&\llangle p,v\rrangle=\int_0^{2\pi} d\phi\, p(\phi) v(\phi).
\end{align}
\end{subequations}
The Lie algebra associated to \eqref{structureCEM} can be found by evaluating considering the infinitesimal adjoint representation \eqref{infadjoint}, which in this case leads to the $\mbms$ algebra \cite{Salgado-Rebolledo:2019kft}. The coadjoint representations of $\Vir$ and $\vir$ follow from these expressions via the definition \eqref{defadcoad}. They are given in terms of the coadjoint representation of $Diff(S^1)$ and its Lie algebra 
\beq\label{CoadDiffS1}
{\rm Ad}^*_{U^{-1}}s= U^\p(\phi)^2 \;s\circ U(\phi) \d \phi^2,
\qquad {\rm ad}^*_v s=- s^\p(\phi) v(\phi) -2 s(\phi) v^\p(\phi).
\eeq
extend to the case of pairs of the form \eqref{pairs1} and
\beq\label{pairs2}
s\rightarrow (s, c_1), \qquad p\rightarrow (p, c_2), \qquad \tilde p\rightarrow (\tilde p, c_3),
\eeq
with pairing
\beq
\llangle S, X\rrangle = \llangle s, x\rrangle +\llangle p, v\rrangle+\llangle \tilde p, \tilde v\rrangle + \sum_{i=1}^3 c_i \,m_i.
\eeq
This leads to
\begin{subequations}\label{CoadVir}
\begin{align}
{\rm Ad}^*_{(U,m_1)} (s,c_1)&= \Big({\rm Ad}^*_{U}s -c_1 \mathcal S(U^{-1}),\,c_1 \Big),
\\
{\rm ad}^*_{(x,m_1)} (s,c_1)&= \Big({\rm ad}^*_x  s +c_1 \mathfrak s(x),\,0 \Big).
\end{align}
\end{subequations}
Using Eqs.~\eqref{AdVir} and \eqref{CoadVir} in the generic expressions \eqref{Adgen} and \eqref{Coadgen}, one finds the adjoint and the coadjoint representation of the Maxwell-BMS$_3$ group \cite{Salgado-Rebolledo:2019kft}. Notice that eliminating $\tilde v$ and $\tilde p$, together with their corresponding central elements, $m_3$ and $c_3$, reproduces the adjoint and the coadjoint representations of the centrally extended BMS$_3$ group \cite{Barnich:2015uva}. As a remark, the group structure \eqref{structureCEM} can be used to define more general Maxwell-like centrally extended groups if the product $UV$ and the cocycle $\mathcal K[U,V]$ are associated to a different centrally extended group.

With these definitions at hand, it is possible to evaluate the geometric action \eqref{genactionMaxwell1} associated with a coadjoint orbit of Maxwell-BMS$_3$. An important detail before doing so is that the Maurer–Cartan form associated with the Virasoro group is not given by $U^{-1}\d U$, which holds for matrix groups. In the case of the diffeomorphism group of the circle, the definition \eqref{MCformsdef} for a diffeomorphism $U(\phi)$ leads to the left-invariant Maurer–Cartan form $\dot U/U^\p$, where the dot denotes the derivative with respect to the worldline parameter, which we identify with the boundary coordinate $u$ of the previous section. In the case of the Virasoro algebra, which includes a central extension defined by the Thurston-Bott cocycle, the usual expression used for the left-invariant Maurer–Cartan form has to be replaced as follows \cite{Alekseev:1988ce,Aratyn:1990dj}
\beq\bal
U^{-1}\d U\rightarrow  \bigg(\frac{\d U}{U^\p}\partial_\phi\;,\;\;
\frac{1}{48\pi}\int d\phi\frac{U^{\p\p}}{U^\p} \left(\frac{\d U}{U^\p}\right)^\p
\bigg).
\eal\eeq
This results from from applying the definition \eqref{MCformsdef} to the $\Vir$ product law \cite{Oblak:2017ect}. For a representative of the form
\beq\label{S0c}
S_{\rm o}=(0,c_1,0,c_2,0,c_3)
\eeq
the geometric action takes the form \cite{Salgado-Rebolledo:2019kft}
\beq\bal
I^{\rm Maxwel-BMS_3}_{\rm GA} &=
\frac{c_1}{48\pi} \int du d\phi  \left( \frac{U^{\p\p}}{U^\p} \right)^\p \frac{\dot U}{U^\p} 
-c_2\int du \,\Big\langle\!\Big\langle
\mathfrak s(\dot U/U^\prime),\, {\rm Ad}_{U^{-1}} v \Big\rangle\!\Big\rangle \\
&-c_3\int du \,\Big\langle\!\Big\langle
\mathfrak s(\dot U/U^\prime),\, {\rm Ad}_{U^{-1}} \tilde v \Big\rangle\!\Big\rangle 
+c_3\int du \,\Big\langle\!\Big\langle
\mathfrak s({\rm Ad}_{U^{-1}} v),\, {\rm Ad}_{U^{-1}} \dot v \Big\rangle\!\Big\rangle .
\eal\eeq
where we have used the following identities 
\begin{subequations}
\begin{align}\label{idSs}
&\mathcal S(U\circ V)= \mathcal S(V)+ {\rm Ad}^*_{V^{-1}} \mathcal S(U),
\\
&\d \mathcal S(U)= {\rm Ad}^*_{U^{-1}} \mathfrak s\Big({\rm Ad}_{U} \frac{\d U}{U^\p}\Big),
\\
&{\rm Ad}^*_U \,\mathfrak s({\rm Ad}_{U^{-1}} v)=\mathfrak s(v)- {\rm ad}^*_v \mathcal S(U^{-1}).
\end{align}
\end{subequations}
To simplify the discussion, we have not included the contribution of the $Diff(S^1)$ representatives $s_{\rm o}$, $p_{\rm o}$, and $\tilde p_{\rm o}$, which can be obtained from \eqref{genactionMaxwell1} using the definitions \eqref{AdDiffS1} and \eqref{CoadDiffS1} corresponding to the diffeomorphism group of the circle without central extension. As shown in \cite{Barnich:2017jgw}, this term can be absorbed into the part of the action associated with a representative of the form \eqref{S0c} by allowing the fields to have nontrivial periodicity conditions.

In order to connect our results with the Maxwellian boundary dual \eqref{maxwellboundarydual}, it is useful to write the geometric action associated with the Maxwell-BMS$_3$ group in the form \eqref{genactionMaxwell2}. By considering the generalization to Virasoro elements given above and evaluating \eqref{genactionMaxwell2} for the orbit representative \eqref{S0c}, we find 
\beq
\bal
&I^{\rm Maxwel-BMS_3}_{\rm GA} =
\frac{c_1}{48\pi} \int du d\phi  \; \frac{\dot U^{\p\p}}{U^\p} 
-c_2\int du \,\Big\langle\!\Big\langle
{\rm Ad}^*_{U^{-1}} \mathfrak s(v) ,\,\frac{\dot U}{U^\prime} \Big\rangle\!\Big\rangle \\
&-c_3\int du \,\Big\langle\!\Big\langle
{\rm Ad}^*_{U^{-1}} \left(\mathfrak 
 s(\tilde v)-\frac12 {\rm ad}_v \mathfrak s(v)\right) ,\,\frac{\dot U}{U^\prime} \Big\rangle\!\Big\rangle  
-\frac{c_3}2 \int du \,\Big\langle\!\Big\langle
\mathfrak s(v),\, \dot v \Big\rangle\!\Big\rangle .
\eal
\eeq
By performing the following field redefinition 
\beq
\varphi = {\rm log} (U^\p), \quad \xi= \frac12 \partial_U (v\circ U)
, \quad \chi=
\frac12 \left(\partial_U (\tilde v\circ U)-\frac12 (v\circ U) \partial^2_U (v\circ U)\right),
\eeq
this action reduces precisely to the kinetic term of the Maxwellian boundary dual \eqref{maxwellboundarydual}
\beq\label{maxwellboundarydualkin}
I^{\rm Maxwel-BMS_3}_{\rm GA}=\int du d\phi\,\left(
\kappa_1\,\varphi^\p\dot\varphi
+\kappa_2\,\xi^\p\dot \varphi
+\kappa_3\,\Big(\chi^\p \dot\varphi+\frac14 \dot\xi \xi^\p
\Big)\right)
\eeq
with the identification $c_i=48\pi \kappa_i$, exactly as in \eqref{candkappa}. As we will see in the following section, in order to complete the equivalence, we need to extended geometric action by including a suitable Hamiltonian, given by a particular conserved charge.

\subsection{Conserved charges and Hamiltonian}
\label{C4}
The conserved charges associated with an infinitesimal global transformation with parameter $X=(\vartheta_1,0,\vartheta_2,0,\vartheta_3,0)$ can be obtained from the definition \eqref{DefQ}, which in this case take the form 
\beq\label{chargemaxwelbms3}
\bal
Q[\vartheta_1,\vartheta_2,\vartheta_3] &=- \Big\langle\!\Big\langle 
{\rm Ad}^*_{(U,m_1,v,m_2,\tilde v, m_3)} (0,c_1,0,c_2,0,c_3), (\vartheta_1,0,\vartheta_2,0,\vartheta_3,0)
\Big\rangle\!\Big\rangle
\\
&= \int d\phi\left( \mathcal J \vartheta_1(\phi) + \mathcal P \vartheta_2(\phi)+ \mathcal Z \vartheta_3(\phi)\right),
\eal
\eeq
where
\begin{subequations}\label{JPZcharges}
\begin{align}
\mathcal J&= c_1 \,\mathcal S(U)+ c_2\, {\rm Ad}^*_{U^{-1}} \mathfrak s(v)
+ c_3\left( {\rm Ad}^*_{U^{-1}} \mathfrak s(\tilde v)
-\frac12 {\rm ad}^*_v \mathfrak s(v)\right),
\\
\mathcal P&= c_2\, \mathcal S(U)+ c_3\, {\rm Ad}^*_{U^{-1}} \mathfrak s(v),
\\
\mathcal Z&= c_3\, \mathcal S(U).
\end{align}
\end{subequations}
The Poisson algebra of the charges is given \eqref{PoissonQ}, and in this case corresponds to the $\mbms$ algebra \eqref{maxwellbms3alg} by construction. Under a global transformation, the dual Lie algebra element entering Eq.~\eqref{chargemaxwelbms3}
\beq
(\mathcal J, c_1, \mathcal P, c_2,\mathcal Z,c_3)=  {\rm Ad}^*_{(U,m_1,v,m_2,\tilde v, m_3)} (0,c_1,0,c_2,0,c_3),
\eeq
transforms under the coadjoint representation of the Maxwell-BMS$_3$ group. This follows from replacing the extension \eqref{pairs2} in the generic form of the coadjoint representation of extended semidirect products \eqref{Coadgen}, and taking the infinitesimal limit. Using \eqref{CoadVir}, this yields
\beq
(\delta\mathcal J, 0, \delta\mathcal P, 0,\delta\mathcal Z,0)=-{\rm ad}^*_{(\vartheta_1,0,\vartheta_2,0,\vartheta_3,0)}(\mathcal J, c_1, \mathcal P, c_2,\mathcal Z,c_3),
\eeq
where
\begin{subequations}\label{deltaJPZ}
\begin{align}
\delta \mathcal J&= \mathcal J^\p \vartheta_1 + 2\mathcal J \vartheta_1^\p - \frac{c_1}{24\pi}\vartheta_1^{\p\p\p}
+ \mathcal P^\p \vartheta_2 + 2\mathcal P \vartheta_2^\p - \frac{c_2}{24\pi}\vartheta_2^{\p\p\p}
+\mathcal Z^\p \vartheta_3 + 2 \mathcal Z \vartheta_3^\p - \frac{c_3}{24\pi} \vartheta_3^{\p\p\p},
\\
\delta \mathcal P&= \mathcal P^\p \vartheta_1 + 2\mathcal P \vartheta_1^\p - \frac{c_2}{24\pi}\vartheta_1^{\p\p\p} + \mathcal Z^\p \vartheta_2 + 2 \mathcal Z \vartheta_2^\p -  \frac{c_3}{24\pi}\vartheta_2^{\p\p\p},
\\
\delta \mathcal Z&= \mathcal Z^\p \vartheta_1 + 2 \mathcal Z \vartheta_1^\p - \frac{c_3}{24\pi} \vartheta_1^{\p\p\p}.
\end{align}
\end{subequations}
This is the explicit form of the infinitesimal coadjoint representation of the Maxwell-BMS$_3$ group, which extends the coadjoint representation of the (centrally extended) $\bms$ algebra \cite{Barnich:2015uva} by including the pair $(\mathcal Z, c_3)$. Defining
\beq
\mathcal Z=\kappa_3 \mathcal M, \qquad 
\mathcal P=\kappa_2 \mathcal M + \kappa _3 \mathcal T, 
\qquad  \mathcal J=\kappa_1 \mathcal M +\kappa_2 \mathcal T+ \kappa_3 \mathcal Y, 
\qquad  \vartheta_i= Y_i ,
\eeq
Eq.~\eqref{deltaJPZ} matches the asymptotic symmetry transformations \eqref{deltaMTY} obtained for the Chern-Simons charges, showing that the three independent charges defined by \eqref{chargemaxwelbms3} close into the $\mbms$ algebra \eqref{maxwellbms3alg}.

It is possible to extend the action \eqref{maxwellboundarydualkin} to the full Maxwellian boundary dual action by introducing a Hamiltonian defined by the charge associated with the translation $X_0=(0,0,-\partial_\phi,0,0,0)$ 
\beq\label{Hmaxwellbms}
\mathcal H= Q[0,-1,0]=-\int d\phi \, \mathcal P= \frac{1}{48\pi} \int d\phi \left( c_2 \varphi^{\p 2} + c_3 \varphi^\p\xi^\p \right).
\eeq
After introducing this Hamiltonian, the modified action becomes precisely the Maxwellian dual model \eqref{maxwellboundarydual} 
\beq
I_{\rm MBD}= I^{\rm Maxwel-BMS_3}_{\rm GA} -\int du \, \mathcal H.
\eeq
Following Ref.~\cite{Barnich:2017jgw}, deforming a geometric action by including a Hamiltonian $\mathcal H = Q_{X_0}$ preserves the symmetries of the original action, provided that the parameters associated with global transformations acquire a fixed time dependence determined by the equation $\dot X(u,\phi) = {\rm ad}_{X_0} X(u,\phi)$. In the case of the $\mbms$ algebra and the Hamiltonian \eqref{Hmaxwellbms}, these conditions lead to the following form for the components of $X(u,\phi)$ in terms of the corresponding time-independent parameters $\vartheta_i(\phi)$ 
\begin{subequations}\label{udependentpar}
\begin{align}
&\vartheta_1(u,\phi) = \vartheta_1(\phi), 
\\
&\vartheta_2(u,\phi) = \vartheta_1(\phi) + u\, \vartheta_1^\p(\phi), 
\\
&\vartheta_3 (u,\phi)= \vartheta_3(\phi) + u\,\vartheta_2^\p(\phi)+\frac{u^2}{2}  \vartheta_1^{\p\p}(\phi),
\end{align}
\end{subequations}
in analogy with \eqref{functionsMaxwell}. When setting $c_3=48\pi \kappa_3=0$, the previous relations reduce to those corresponding to the geometric action on coadjoint orbits of the BMS$_3$ group \cite{Barnich:2013yka}.

It is important to remark that, in complete analogy with the analysis done in this section, the kinetic term of Maxwell-invariant chiral WZW model \eqref{MaxwellWZWaction} can be shown to be equivalent to the geometric action \eqref{genactionMaxwell2}  on coadjoint orbits of the Maxwell-Kac-Moody group, which was given in the alternative the form \eqref{geometricaction1} in \cite{Salgado-Rebolledo:2019kft}. One can see this by replacing the extension \eqref{pairs1} with
\beq
\Lambda \rightarrow (\lambda, m_1),  \qquad
 \qquad a\rightarrow (\beta, m_2), \qquad \tilde a\rightarrow (\gamma, m_3),
\eeq
and using the definitions \eqref{AdVir} and \eqref{CoadVir} adapted to the $\SL$ loop group. This requires to consider the definitions \eqref{AdMatrix} for angle-dependent $\SL$ and $\sl$ elements with a pairing given by the integral on $S^1$ of the $\SL$ Killing form, and to define the central extension through the Souriau cocyle 
\beq
\mathcal S(\lambda)=\frac{1}{2\pi} \lambda^{-1}\lambda^\p, \qquad 
\mathfrak s(\beta)= \frac{1}{2\pi} \beta^\p.
\eeq
This leads precisely to the Maxwell-WZW model \eqref{MaxwellWZWaction} with $c_i=4\pi\kappa_i$. after supplementing centrally extended version of the action \eqref{genactionMaxwell2} with the Hamiltonian
\beq\label{HamiltonianWZW}
\mathcal H= \frac{\pi}{c_3}{\rm Tr}
\left(
2\mathcal P \mathcal Z
-\frac{c_2}{c_3} \mathcal Z^2
\right).
\eeq
The conserved charges that follow from \eqref{JPZcharges} in this case reproduce precisely the time independent part of the expression for the charges given in \eqref{ChargesMKM} and therefore satisfy the Kac-Moody algebra \eqref{MaxwellKM}. The full expression of the charges is recovered after introducing the Hamiltonian \eqref{HamiltonianWZW}, where the relation $\dot X(u,\phi) = {\rm ad}_{X_0} X(u,\phi)$ for a $u$-independent $X_0=(\vartheta_1,\vartheta_2,\vartheta_3)$ modifies the gauge parameters precisely as in \eqref{udependentpar}.

\subsection{Carrollian expansion of the AdS$_3$ boundary dual}
\label{C3}
In this section, we will show how the Maxwellian boundary dual theory \eqref{maxwellboundarydual} can be obtained from a Carrollian expansion of the AdS$_3$ boundary dual. This is a natural generalization of the fact that the flat Liouville model, known as the boundary dual to asymptotically flat three-dimensional gravity \cite{Barnich:2013yka}, arises as a Carroll limit of Liouville theory, the latter being the corresponding dual in the AdS$_3$ case \cite{Coussaert:1995zp}. In order to see this, let us first recall the flat limit of Liouville theory, whose action is given by
\beq
I_{\rm Liouville}= \frac\kappa 2 \int d^2x \left(\partial_\alpha \Psi \partial^\alpha \Psi-\mu e^{ \Psi}\right),
\eeq
where $\Psi$ is the Liouville field and indices are raised and lowered with the two-dimensional Minkowski metric $\eta_{\alpha\beta}={\rm diag}(-1,1)=\eta^{\alpha\beta}$. We use coordinates $x^\alpha =(c u,\phi)$, where $c$ denotes the speed of light. By redefining the constant $\kappa$ as
\beq
\kappa= \frac{1}{c\nu^2},
\eeq
to absorb the factor of $c$ from the integration measure, and rescaling the Liouville field as $\Psi = \nu \Phi$ with $\nu$ an arbitrary constant, we obtain \cite{Barnich:2012rz}
\beq
I_{\rm Liouville}= \frac12 \int dud\phi \left(\frac1{c^2}\dot\Phi^2 -\Phi^{\p 2}-\frac{\mu}{\nu^2} e^{ \nu\Phi}\right).
\eeq
Following \cite{Henneaux:2021yzg}, we define the Carrollian limits of the action in the Hamiltonian formulation by introducing the canonical momentum
\beq
\Pi=\frac1{c^2}\dot\Phi,
\eeq
which allows us to express the action in Hamiltonian form 
\beq
I_{\rm Liouville}= \int dud\phi \left(\Pi\dot\Phi -\frac{c^2}2 \Pi^2 -\frac12 \Phi^{\p 2}-\frac{\mu}{2\nu^2} e^{ \nu\Phi}\right).
\eeq
The magnetic Carrollian limit arises by taking $c\rightarrow0$ directly in the Hamiltonian action, yielding the flat Liouville theory \cite{Barnich:2012rz} 
\beq
I_{\rm Liouville}^{\rm MC}= \frac12 \int dud\phi \left(\Pi\dot\Phi-\frac12 \Phi^{\p 2}-\frac{\mu}{2\nu^2} e^{ \nu\Phi}\right).
\eeq
The electric Carrollian limit, on the other hand, follows from the rescaling 
\beq
\Pi\rightarrow c^{-1}\Pi,\quad \Phi\rightarrow c\Phi,\qquad
\nu\rightarrow c^{-1} \nu,\qquad
\mu\rightarrow c^2 \mu,
\eeq
which leads to the action 
\beq
I_{\rm Liouville}^{\rm EC}= \int dud\phi \left(\Pi\dot\Phi -\Pi^2 -\frac{\mu}{2\nu^2} e^{ \nu\Phi}\right).
\eeq
The electric limit corresponds to a different flat limit of Liouville theory, also considered in Ref.~\cite{Barnich:2012rz}.

The equivalence between the magnetic Carrollian limit of Liouville theory and the flat Liouville theory emerging from asymptotically flat three-dimensional gravity \cite{Barnich:2013yka} can be established through the field redefinition 
\beq\label{redefflatLiouville}
\bal
\nu \Phi &= 2\varphi-2\,{\rm ln}\sigma+{\rm ln}\frac{4}{\mu},
\\
\nu \Pi &= \xi^\p-({\rm ln}\sigma)^\p\xi, \qquad \sigma^\p=e^\varphi,
\eal
\eeq
which reproduces the $\kappa_2$-term in \eqref{maxwellboundarydual} with $\nu^2=1/\kappa_2$. However, the $\kappa_1$-term cannot be recovered by considering purely Einstein gravity in 2+1 dimensions, since it originates from the so-called exotic gravitational term \cite{Witten:1988hc}. Starting instead from an AdS$_3$-invariant Chern-Simons theory with the most general non-degenerate bilinear form, the action can be written in terms of two $\SL$ Chern-Simons terms
\beq
I[\mathcal A_\pm]=\int d^3 x\, {\rm Tr}\left[\mathcal A_\pm \wedge \d \mathcal A_\pm +\frac23 \mathcal A_\pm\wedge \mathcal A_\pm\wedge \mathcal A_\pm\right]
\eeq
as
\beq
I_{\rm AdS}=(\mu_J + \mu_P) I[\mathcal A_+] + (\mu_J - \mu_P) I[\mathcal A_-].
\eeq
Einstein gravity is recovered when setting $\mu_J=0$, whereas the exotic term corresponds to $\mu_P=0$. After imposing the Hamiltonian Chern-Simons constraint and applying the Drinfeld-Sokolov reduction associated with Brown-Henneaux boundary conditions, the boundary theory reduces to a sum of two chiral bosons 
\beq\label{chiralbosons}
I_{\rm AdS}=\frac{\mu_J+\mu_P}{2}\int c\,du d\phi\;\left(
\frac{1}{c}\dot\zeta_+ \zeta_+^\p - \zeta_+^{\p2}
\right)
+ 
\frac{\mu_J-\mu_P}{2}\int c\,du d\phi\;\left(\frac{1}{c}\dot\zeta_- \zeta_-^\p + \zeta_-^{\p2}
\right).
\eeq
For $\mu_J=0$, the standard Einstein gravity result is recovered \cite{Coussaert:1995zp}, which can be identified with Liouville theory after the field redefinitions \cite{Gervais:1982nw,Henneaux:1999ib} 
\beq\label{redefLiouville}
\bal
\nu\Psi&=\zeta_+ +\zeta_- -2\text{ln}(\sigma_+ +\sigma_-)+{\rm ln}\frac{16}{\mu},
\\
c\nu\Pi&=\zeta_+^\p -\zeta_+^\p-2\frac{\sigma_+^\p -\sigma_-^\p}{\sigma_+ +\sigma_-}, \qquad \sigma_\pm^\p=e^{\zeta_\pm}.
\eal
\eeq
One can see that a Carrollian limit connecting \eqref{redefLiouville} with the ``flat'' redefinition \eqref{redefflatLiouville} is achieved by setting $\zeta_\pm=\varphi\pm\frac{c}{2}\xi$ and taking $c\rightarrow 0$. This motivates the following Carrollian expansion for the AdS$_3$ boundary dual field theory 
\beq\label{carrollexp}
\zeta_\pm=\sum_{m=0}^\infty (\pm c)^m  \zeta_m.
\eeq
Substituting this into the action \eqref{chiralbosons}, we obtain up to total derivatives
\beq\label{IAdSexpansion}
\bal
I_{\rm AdS}&=\int dud\phi\Bigg(
\mu_J\sum_{m=0}^\infty c^{2m}\left(
\sum_{n=0}^{2m} \dot\zeta_n \zeta_{2m-n}^\p -\sum_{n=0}^{2m-1} \zeta_n^\p\zeta_{2m-n-1}^\p
\right)
\\
&+\mu_P\sum_{m=0}^\infty c^{2m+1}\left(
\sum_{n=0}^{2m+1} \dot\zeta_n \zeta_{2m-n+1}^\p -\sum_{n=0}^{2m} \zeta_n^\p\zeta_{2m-n}^\p
\right)
\Bigg),
\eal
\eeq
and considering this Carrollian expansion up to order $c^2$ we find
\beq
I_{\rm AdS}= \int du d\phi \, \left(
\mu_J \,\zeta_0^\p \dot{\zeta_0}
+ c\,\mu_P\, \big(2\zeta_1^\p \dot{\zeta_0} -  \zeta_0^\p  \zeta_0^\p\big)
+ c^2\mu_J \,\Big( 2\zeta_2^\p \dot{ \zeta_0} + \dot{ \zeta_1}  \zeta_1^\p - 2 \zeta_1^\p \zeta_0^\p \Big)
+ O(c^3)\right).
\eeq
Thus, we recover the Maxwellian boundary dual model \eqref{maxwellboundarydual} with the following identifications
\beq
\bal
&\kappa_1=\mu_J,&\qquad& \kappa_2 =c\,\mu_P,&\qquad& \kappa_3= c^2\mu_J
\\
&\varphi=\zeta_0,&\qquad& \xi=2\zeta_1,&\qquad& \chi=2\zeta_2.
\eal
\eeq
Notice that the extension of the flat Liouville theory given in Refs. \cite{Barnich:2015sca,Barnich:2017jgw}, which also follows from the action \eqref{maxwellboundarydual} for $\kappa_2=0$, can be obtained as the Carroll limit $c\rightarrow 0$ of \eqref{IAdSexpansion} after rescaling $\mu_P \rightarrow \mu_P / c$.

\section{Conclusions and outlook}
\label{D}

In this paper, we have analyzed the boundary dynamics of Maxwell$_3$-invariant Chern-Simons gravity in 2+1 dimensions using standard Hamiltonian reduction techniques. Solving the constraints in the gravitational action leads to a chiral WZW model invariant under the loop extension of the Maxwell group, with conserved charges realizing the corresponding Kac-Moody symmetry. By imposing boundary conditions that naturally generalize those associated with asymptotically flat spacetimes in the BMS gauge \cite{Concha:2018zeb}, the WZW model is further reduced to a Maxwellian extension of the flat Liouville theory of Barnich, Gomberoff, and Gonz\'alez \cite{Barnich:2012rz}.

Alternatively, we have shown that the boundary dual to Maxwell$_3$-invariant Chern-Simons gravity is equivalent to the geometric action on coadjoint orbits of the Maxwell-BMS$_3$ group, for a specific class of orbit representatives and upon inclusion of a suitable Hamiltonian. This Hamiltonian preserves the symmetries of the original action, provided the global symmetry parameters acquire a specific fixed dependence on the temporal coordinate. The conserved charges of the model, computed via the coadjoint orbit method, naturally realize the $\mbms$ algebra.

By identifying the flat Liouville theory known from asymptotically flat three-dimensional gravity as an electric Carroll limit, we demonstrated that the Maxwellian boundary dual presented here emerges from a Carrollian expansion of the dual field theory associated with AdS$_3$ Chern-Simons gravity. This indicates that the contributions introduced by the Maxwell extension can be interpreted as a post-Carrollian correction.

A natural question arising from these results is whether such an expansion can be extended to higher orders in $c$ at the boundary, and what the corresponding gravitational dual would be in that case. Another interesting direction is the generalization to four dimensions, which could relate Maxwell gravity \cite{deAzcarraga:2010sw} to a Carrollian expansion of a relativistic CFT. We expect to explore these questions in future work.

A compelling generalization of our results involves the higher-spin extension of the Maxwell-BMS$_3$ symmetry, namely a Maxwell-like $\mathcal{W}_N$ algebra. Such symmetry was first introduced in \cite{Salgado-Rebolledo:2019kft} and later explicitly constructed in \cite{Concha:2024rac} as the asymptotic symmetry associated with the Maxwell extension of a Chern-Simons higher-spin gravity theory \cite{Caroca:2017izc}. Following the methodology developed in this work, the Maxwellian boundary theory of \cite{Concha:2024rac} could be extended to the higher-spin case, leading to a Maxwellian extension of the flat Toda theory \cite{Gonzalez:2014tba}, which arises as the boundary dual of asymptotically flat higher-spin Chern-Simons gravity \cite{Afshar:2013vka,Gonzalez:2013oaa}. Since flat $\mathcal{W}_N$ algebras arise from ultra-relativistic limits of $\mathcal{W}_N \otimes \mathcal{W}_N$ algebras \cite{Campoleoni:2016vsh}, the resulting Maxwellian extension is likely related to a Carrollian expansion of Toda theory.

Finally, another interesting Maxwell-like symmetry studied in the literature is the so-called AdS-Lorentz algebra \cite{Soroka:2004fj,Gomis:2009dm}, which defines a semisimple extension of the Poincar\'e symmetry, and extends the Maxwell algebra by incorporating a cosmological constant term. In \cite{Concha:2018jjj}, it was shown that the asymptotic symmetry of a Chern-Simons gravity theory with AdS-Lorentz gauge symmetry corresponds to three copies of the Virasoro algebra. In the vanishing cosmological constant limit, this algebra reduces to the $\mbms$ algebra \eqref{maxwellbms3alg}. This suggests that such a Chern-Simons theory admits a boundary dual described by three chiral boson actions, whose flat limit reproduces the Maxwellian boundary dual considered here.

\section*{Acknowledgements}

We thank Srinath Bulusu, Florian Ecker, Daniel Grumiller, Arash Ranjbar and Misao Sasaki for very interesting comments and discussions. P.S.-R.~has been supported by a Young Scientist Training Program (YST) fellowship at the Asia Pacific Center for Theoretical Physics (APCTP) through the Science and Technology Promotion Fund and the Lottery Fund of the Korean Government. P.S.-R. has also been supported by the Korean local governments in Gyeongsangbuk-do Province and Pohang City.

\providecommand{\href}[2]{#2}\begingroup\raggedright\endgroup

\end{document}